\documentclass[%
reprint,
 amsmath,amssymb,
 aps,
prstab,
floatfix,
]{revtex4-1}

\usepackage{graphicx}% Include figure files
\usepackage{dcolumn}% Align table columns on decimal point
\usepackage{bm}% bold math
\usepackage[hidelinks]{hyperref}% add hypertext capabilities
\usepackage{tikz} % allowed?
\usepackage{url}
\usepackage{units}

\usepackage[colorinlistoftodos,prependcaption,textsize=tiny]{todonotes}

\usepackage[utf8]{inputenc}

\begin{document}

\preprint{APS/123-QED}

\title{Synchronous Detection of Longitudinal and Transverse Bunch Signals at a Storage Ring}
\author{Benjamin Kehrer}
\email{benjamin.kehrer@kit.edu}
\author{Miriam Brosi}
\author{Johannes L. Steinmann}

\author{Edmund Blomley}
\author{Erik Bründermann}
\author{Michele Caselle}
\author{Stefan Funkner}
\author{Nicole Hiller}
\thanks{now at PSI, Villigen, Switzerland}

\author{Michael J. Nasse}
\author{Gudrun Niehues}
\author{Lorenzo Rota}
\thanks{now at SLAC, Menlo Park, CA, USA}
\author{Manuel Schedler}
\thanks{now at VARIAN PT,  Troisdorf, Germany}

\author{Patrik Schönfeldt}
\thanks{now at DLR-VE, Oldenburg, Germany}
\author{Marcel Schuh}
\author{Paul Schütze}
\thanks{now at DESY,  Hamburg, Germany}

\author{Marc Weber}
\author{Anke-Susanne Müller}

\affiliation{KIT, Karlsruhe, Germany}

\date{\today}

\begin{abstract}

To understand and control dynamics in the longitudinal phase space, time-resolved measurements of different bunch parameters are required. For a reconstruction of this phase space, the detector systems have to be synchronized. This reconstruction can be used for example for studies of the micro-bunching instability which occurs if the interaction of the bunch with its own radiation leads to the formation of sub-structures on the longitudinal bunch profile. These sub-structures can grow rapidly -- leading to a sawtooth-like behaviour of the bunch.
At KARA, we use a fast-gated intensified camera for energy spread studies, Schottky diodes for coherent synchrotron radiation studies as well as electro-optical spectral decoding for longitudinal bunch profile measurements. For a synchronization, a synchronization scheme is used which compensates for hardware delays. In this paper, the different experimental setups and their synchronization are discussed and first results of synchronous measurements presented. 
 
\end{abstract}

\pacs{Valid PACS appear here}
\maketitle

\section{Introduction}
The investigation of the dynamics of short bunches in a storage ring sets stringent requirements for the diagnostics: Response time and repetition rate of the different detector systems must be sufficient to study a single bunch in a multi-bunch environment on a single-turn basis. The required time scales are given by the RF frequency and the revolution time. In addition, the different detector systems must be synchronized to study simultaneous changes for example in the bunch shape and in the emitted synchrotron radiation. 

\section{Micro-bunching instability}

Under certain conditions, periodic bursts of coherent synchrotron radiation (CSR) were recorded at different facilities (e.g. NSLS VUV \cite{CARR2001387}, SURF II \cite{PhysRevSTAB.4.054401}, ALS \cite{PhysRevLett.89.224801}, BESSY-II \cite{abo2003coherent}, Diamond \cite{Shields_Diamond}, SOLEIL \cite{Evian_MBI_EPL}). Theoretical studies of this phenomena gave hints to a self-interaction of the bunch with its own wake-field \cite{PhysRevSTAB.5.054402}, which can lead to the occurrence of sub-structures in the longitudinal phase space. The rising amplitude of these sub-structures can be explained by the CSR impedance \cite{PhysRevLett.89.224802}. The rapid increase of the amplitude is coupled to an overall increase in projected bunch size. At one point, diffusion and radiation damping outweigh the growth of the sub-structures and they start to dissolve and damp down, leading to a reduction of the bunch size. The reduced bunch size leads again to a stronger wake-field and subsequently the next CSR instability cycle starts over. Thus, the bunch size is modulated with a repetitive sawtooth pattern in time \cite{Warnock:2006qa}. Due to the changing bunch profile, the CSR is emitted in bursts and the behavior is referred to as \textit{bursting}. 

As the longitudinal phase space is spanned by time and energy, it can be studied by time-resolved measurements of the longitudinal bunch profile and energy spread. The CSR intensity depends on the longitudinal bunch profile, so it can also be used to probe the dynamics. A monitoring of the longitudinal phase space leads to a deeper understanding of the beam dynamics, especially above the bursting threshold and could potentially lead to methods for control. Such a reconstruction requires synchronous measurements of the different bunch properties.

At KARA, we use different diagnostics tools for theses measurements which are integrated in a hardware synchronization system to enable simultaneous data acquisition. 

\section{KARA}

KARA (Karlsruhe Research Accelerator), the \unit[2.5]{GeV} KIT storage ring, can be operated in different modes including a short bunch mode. In this mode, the momentum-compaction factor $\alpha_c$ is lowered by changing the magnet optics \cite{muller2012beam} to study the micro-bunching instability \cite{PhysRevAccelBeams.19.110701, steinmann2017continuous}. 

At KARA, the revolution time is \unit[368]{ns} and the bunch spacing of \unit[2]{ns} is defined by the radio frequency (RF) of \unit[500]{MHz}. One single measurement should take at most \unit[2]{ns}, so that only the signal of one bunch is detected. 
As part of the bunch detection setup, several systems have been installed and commissioned: A fast-gated intensified camera (FGC) to measure the horizontal bunch position and size, fast THz detectors to detect the coherent synchrotron radiation (CSR) as well as an electro-optical bunch profile monitor to determine the longitudinal bunch profile \cite{fisher2006turn, hubers2005time, Wilke_EOSD_SingleShot_PRL}.

\section{Horizontal bunch size measurements \label{sec:FGC}}

Measurements of the horizontal bunch size $\sigma_x$ allow studies of the energy spread $\sigma_\delta$ as they are related by  \cite{wiedemann2007particle}
\begin{equation*}
\sigma_x = \sqrt{\beta_x \cdot \epsilon_x + \left( D_x \cdot \sigma_\delta \right)^2}
\end{equation*}

in a dispersive section. 
Therefore, a time-resolved measurement of the horizontal bunch size reveals changes in energy spread. We use a fast-gated intensified camera (FGC). It is located at the Visible Light Diagnostics (VLD) port at KARA which uses incoherent bending radiation from a 5$^\circ$ port of  a dipole magnet \cite{KehrerIPAC2015}. The FGC setup consists of a camera and a fast-rotating mirror which sweeps the incoming light over the entrance aperture of the FGC during the image acquisition. In combination with a fast switching of the image intensifier, this allows single turn images of a bunch, even in a multi-bunch fill \cite{SchuetzeIPAC2015, Schuetze_master}.
The resolution of the setup is confined by a rectangular crotch absorber with a horizontal width of \unit[20]{mm} which is located \unit[1600]{mm} downstream of the radiation source point \cite{HillerIPAC2011}. Taking diffraction effects into account, the resolution is estimated to be \unit[77]{$\mu$m} using a wavelength of \unit[400]{nm} \cite[Eq. 30]{hofmann1982optical}.

\begin{figure}[t]
\includegraphics[width=\columnwidth]{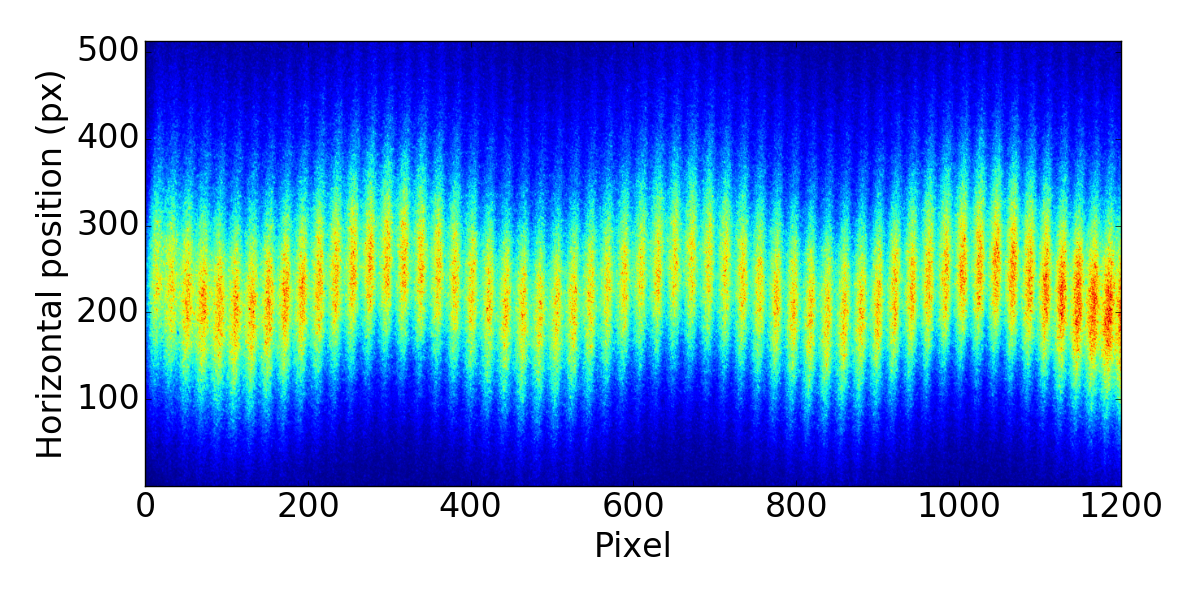}
\caption{Raw image taken by the fast-gated intensified camera (FGC). It shows 58 projections of the same bunch for every 14th turn. The optical path of the setup rotates the spots by \unit[90]{$^\circ$} so their horizontal axis is displayed on the vertical axis of the image.}
\label{fig:FGC_sample_Picture}  
\end{figure}

Fig.~\ref{fig:FGC_sample_Picture} shows a raw image recorded with the FGC. 

\begin{figure}[t]
\includegraphics[width = \columnwidth]{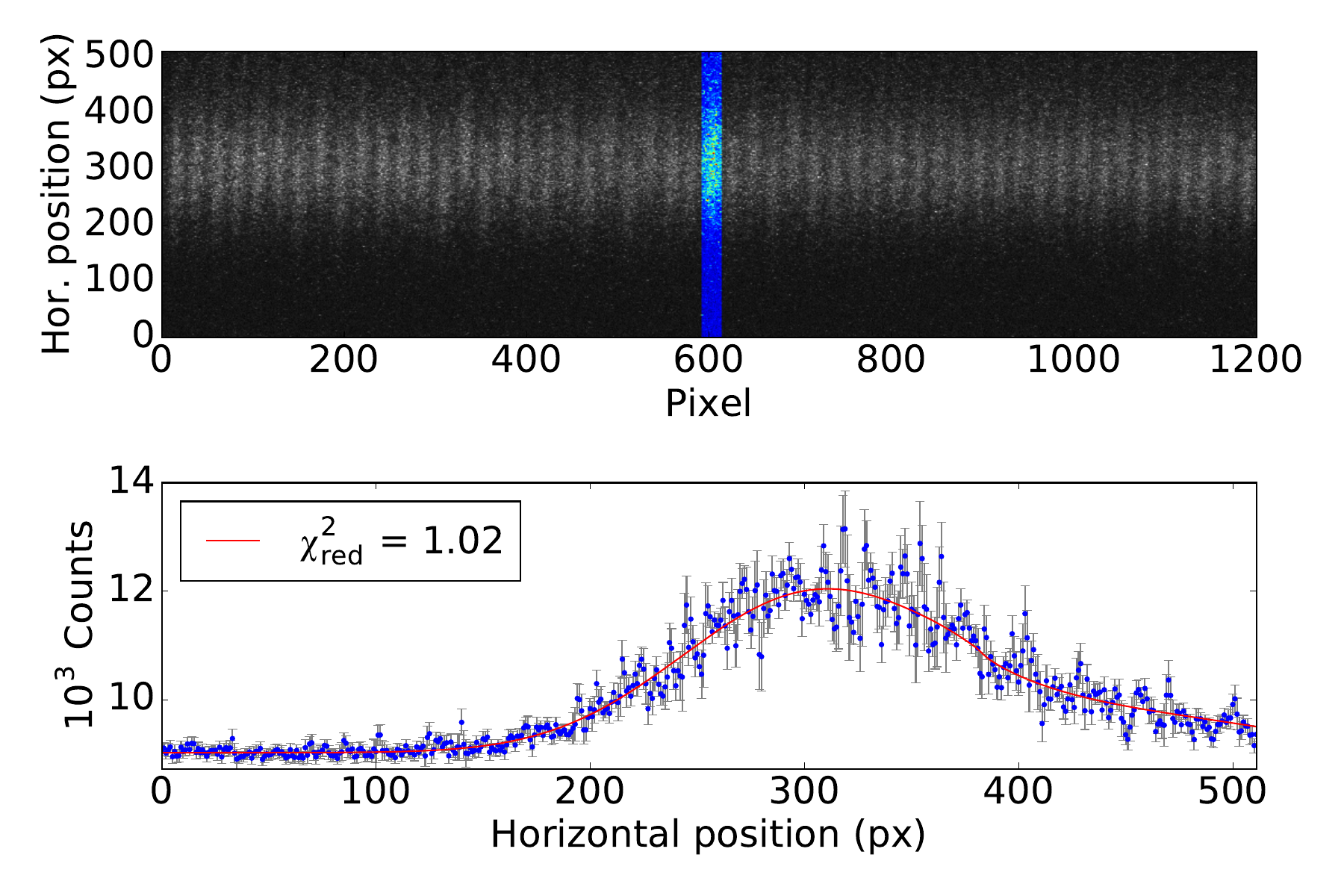}
\caption{FGC raw image in grey scale with the analysis range for one spot highlighted in color is shown in the top panel. The corresponding profile and the fit is plotted in the bottom panel.}
\label{fig:FGC_FBSF_ExampleSpot}
\end{figure}

Here, the machine was operated without an active feedback system, therefore the bunch is undergoing a synchrotron oscillation ($f_s \approx$ \unit[11.24]{kHz}). 
The individual spots on this image are the convolution of the charge distribution with the so-called \textit{filament beam spread function} (FBSF). The FBSF can be seen as an extension of the point spread function for a moving point-like source, in this case a moving single electron \cite{oleg2006electron}. We use the software OpTalix \footnote{OpTalix Pro 8.82, www.optenso.com} to simulate the optical system and the imaging process. Instead of deconvolving the FBSF from the data and fitting a Gaussian curve, we fit a convolution of the FBSF with a Gaussian curve. Compared to the deconvolution this has the advantage that no estimation of the signal-to-noise ratio is required, in addition, this method is fast and robust. For one spot of an FGC raw image, this analysis is illustrated in Fig.~\ref{fig:FGC_FBSF_ExampleSpot}. 

It can be seen that the fit reproduces the distorted shape of the profile very well, enabling a determination of the horizontal bunch size with single turn resolution. 

\begin{figure}[b]
\centering
\includegraphics[width = \columnwidth]{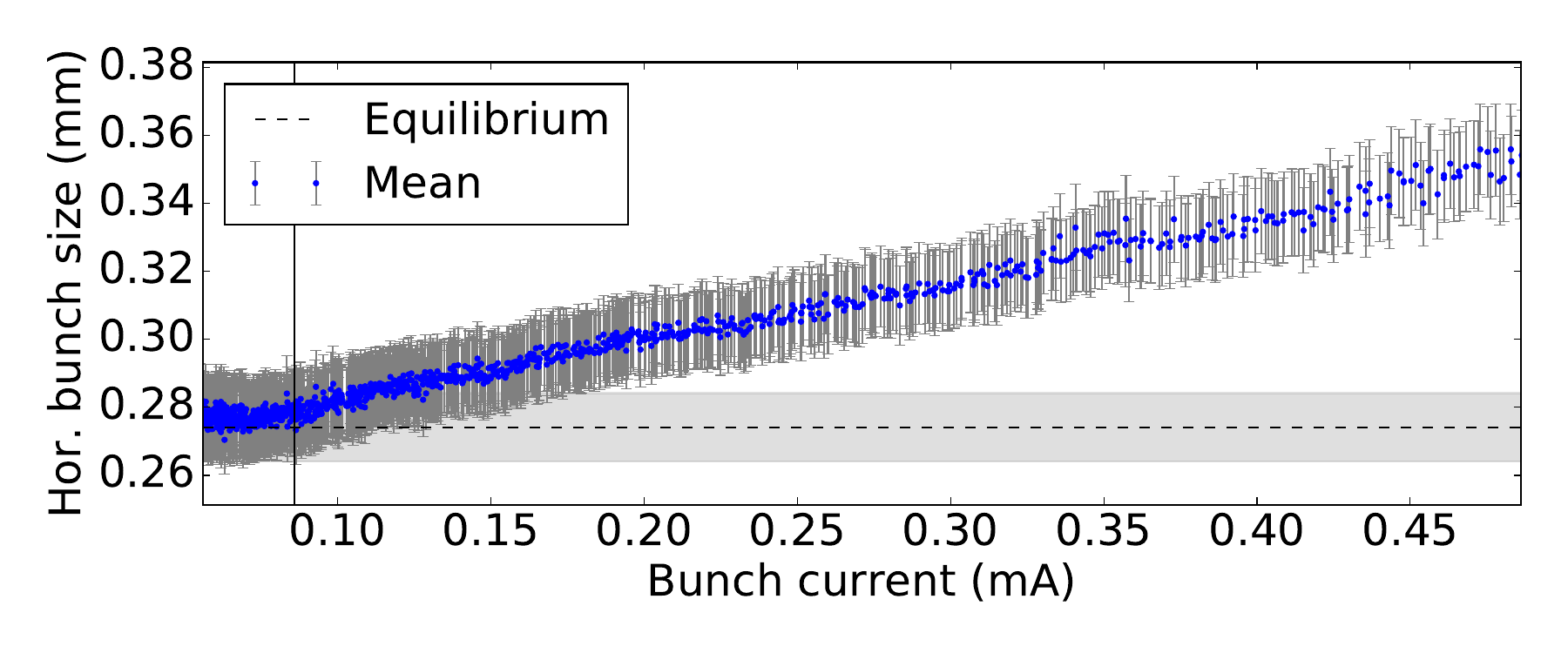}
\caption{Mean horizontal bunch size over bunch current. The dashed line and its grey error bar are depicting the simulated equilibrium bunch size at the VLD source point, while the vertical black line illustrates the calculated bursting threshold.}
\label{fig:f05896_2016-04-06_FGC_EquilibriumBunchSize}
\end{figure}

Experimental studies showed that the horizontal bunch size remains constant below the bursting threshold, while it starts to increase for currents above the threshold \cite{Bane2005}.
For the equilibrium case below the bursting threshold, the measured bunch size is in good agreement with simulated values using  the accelerator toolbox for Matlab (AT) \cite{nash2015new} and the lattice model of KARA, see Fig.~\ref{fig:f05896_2016-04-06_FGC_EquilibriumBunchSize}. For the AT model, LOCO fits \cite{safranek2009linear} are used to determine the quadrupole strengths from measurements of the orbit response matrix and the dispersion.

A quantitative determination of the energy spread is not possible due to the unknown contribution of the horizontal emittance to the horizontal bunch size, therefore we are limited to qualitative studies with the horizontal bunch size taken as a measure for the energy spread. Besides the studies of the micro-bunching instability, the system can be used to study different kinds of instabilities, e.g. during the injection or on the energy ramp.    

\section{Coherent synchrotron radiation}

The investigation of the CSR dynamics includes THz detectors that are fast enough to resolve the emitted synchrotron radiation for each bunch in a multi-bunch environment.   
We use commercially available room temperature zero-bias Schottky diode detectors with different sensitivity ranges \cite{PhysRevAccelBeams.19.110701, steinmann2017continuous}. 
To read out these detectors we use the data acquisition (DAQ) system KAPTURE \cite{caselle2014ultra, Caselle2017KAPTURE}. KAPTURE is a fast FPGA-based system with four input channels that can be used to sample the signal of one detector with four sample points per bunch and turn. Alternatively, it is possible to sample up to four detectors in parallel with a bunch-by-bunch sampling rate of \unit[500]{MHz}. For the measurements presented here, we configured KAPTURE to digitize the peak intensity of multiple different detectors simultaneously. 

\begin{figure}[b]
\includegraphics[width=\columnwidth]{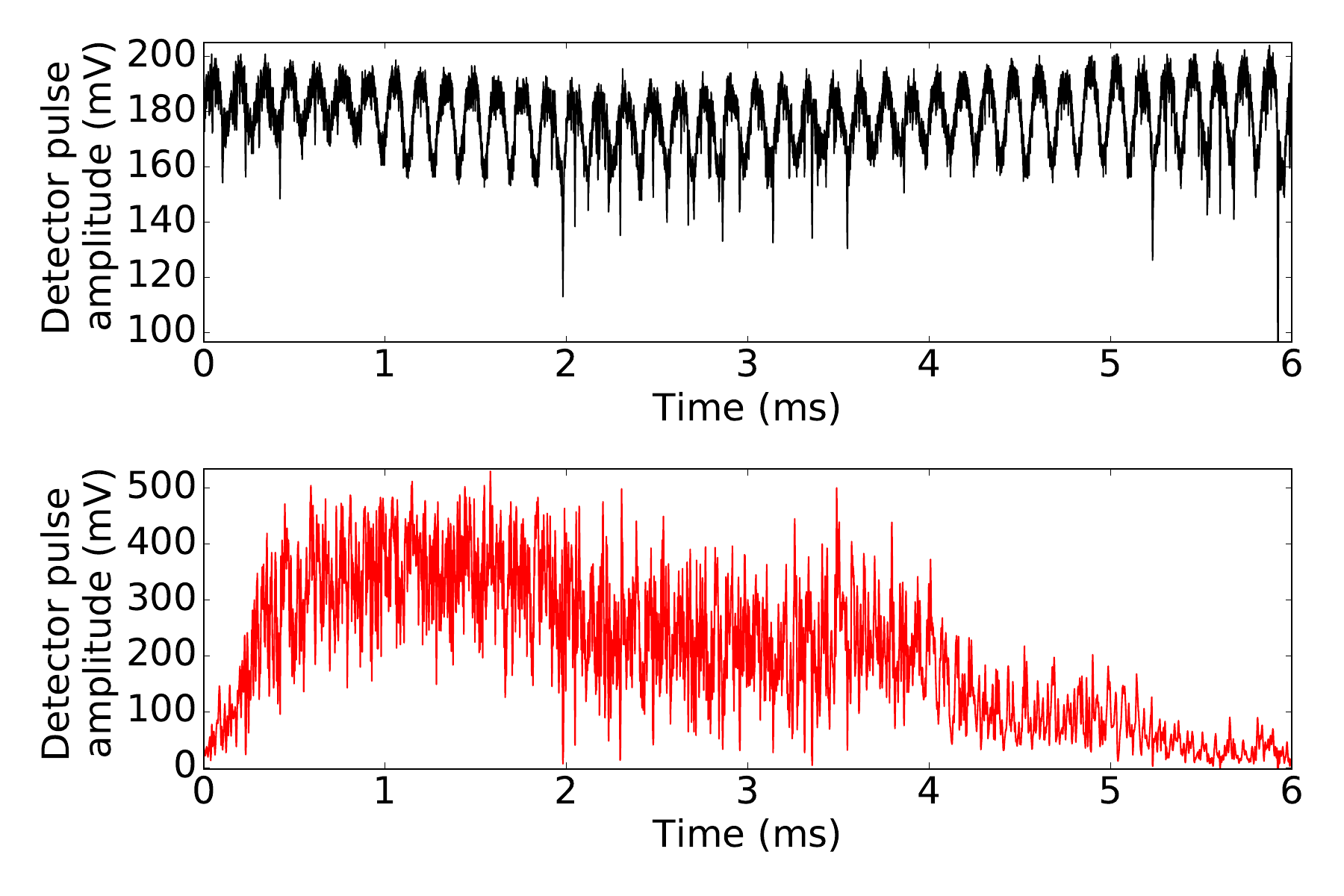}
\caption{Data recorded by KAPTURE using two different detectors in parallel. The top panel shows the signal from an APD in the visible range while the bottom panel shows the turn-by-turn THz signal recorded using a narrow-band Schottky detector \cite{VDI}. The THz signal has the sawtooth-like pattern induced by the bursting and the APD signal is modulated by the synchrotron frequency.}
\label{fig:f05794_2015-11-06T23h37m57s_KAPTUREsignal}
\end{figure}

Figure~\ref{fig:f05794_2015-11-06T23h37m57s_KAPTUREsignal} illustrates this multi-detector operation mode. It shows the signals recorded by an avalanche photo diode (APD) sensitive to incoherent synchrotron radiation in the visible range and a Schottky diode sensitive in the THz range. The APD is insensitive to arrival time oscillations caused by synchrotron oscillation, as the duration of the APD pulse is longer than the amplitude of the synchrotron oscillation and KAPTURE samples at a fixed phase relative to the RF. Thus, the observed oscillation of the APD signal is due to an intensity fluctuation. The bunch oscillates horizontally around the focal point of the optics and the limited aperture of the beam line optics transfers this into an intensity oscillation.
This signal will play a vital role for the calibration of the synchronization in Sec.~\ref{sec:synchronization}.

The combination of KAPTURE and Schottky diodes is used for detailed and systematic studies of the CSR emission during the micro-bunching instability. The analysis of the fluctuation of the CSR intensity allows for example a fast and precise mapping of the bursting threshold \cite{PhysRevAccelBeams.19.110701}. Using four Schottky diodes with different frequency bands we can configure the KAPTURE system into a bunch-by-bunch spectrometer for CSR studies \cite{steinmann2017continuous}.

\section{Longitudinal bunch profile}

The technique of electro-optical spectral decoding (EOSD) \cite{ZhangFirstEOSD} is used to determine the longitudinal bunch profile and arrival time \cite{roussel2016electro}. This is done by inserting an electro-optical crystal into the vacuum pipe to sample the  electric near-field of the bunch \cite{hillerIPAC2014}. This field turns the crystal birefringent. A laser pulse, chirped on the ps-time scale, is directed through the crystal and the induced birefringence turns the initial linear polarization into an elliptical. Thus the bunch profile is imprinted into the polarization of the laser pulse. A crossed polarizer transfers this into an intensity modulation and, due to the unique, ideally linear, correlation between time and wavelength in the chirped laser pulse, the bunch profile can be determined by recording the modulated laser pulse spectrum. Measuring in the near field has the advantage that there is no frequency cut-off of the electric field by the beam-line.

\begin{figure}[h]
\includegraphics[width=\columnwidth]{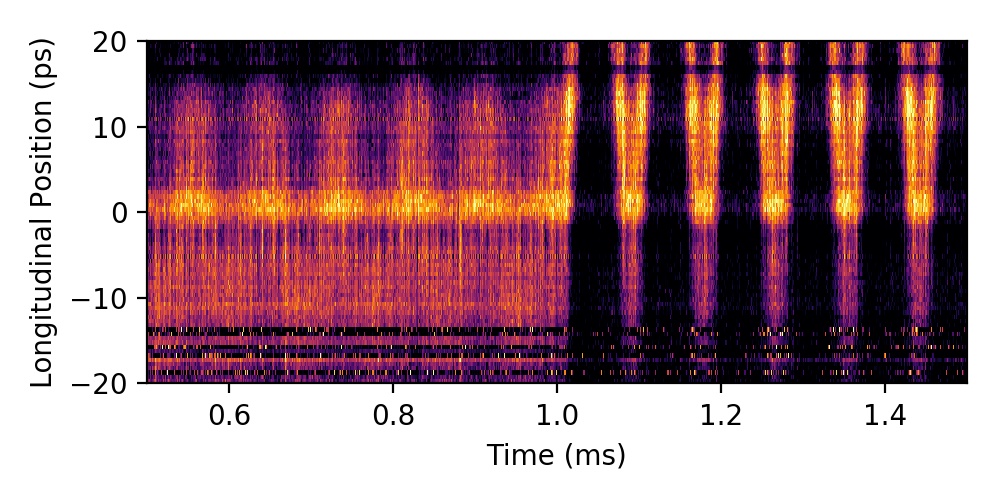}
\caption{Longitudinal bunch profiles recorded with the KALYPSO-based electro-optical spectral decoding setup. Each image column corresponds to one bunch profile. Synchrotron oscillations can be seen during the entire time range.
The abrupt change of its amplitude at \unit[1]{ms} is due to an RF phase step that leads to an oscillation around the new phase. The upper part of this oscillation is not visible due to the finite size of the line array and the length of the chirped laser pulse chosen to achieve a better temporal resolution.}
\label{fig:f05741_2015-10-14T16h36m35s484358091_EOSD_Si2-2400-3900-plot}
\end{figure}

The spectrometer used for this measurement is based on the KALYPSO system (KArlsruhe Linear arraY detector for MHz-rePetition rate SpectrOscopy \cite{ROTA_IBIC_16, CaselleIBIC2017}). KALYPSO consists of a 256-pixel line array that can be read out with a turn-by-turn frame rate of \unit[2.7]{MHz}. 

An example recorded by this system is shown in Fig.~\ref{fig:f05741_2015-10-14T16h36m35s484358091_EOSD_Si2-2400-3900-plot}. It shows color-coded longitudinal bunch profiles (in units of ps) as a function of time. These measurements were taken with an early development version of the KALYPSO system, which manifests in pixel errors and other artifacts. They are visible for example as dark horizontal lines between \unit[-10 and -20]{ps}. At the beginning, the bunch is undergoing a synchrotron oscillation as also here the feedback system was switched off. At \unit[1]{ms}, this oscillation abruptly changes its amplitude due to a triggered step in the RF phase at this point in time. For these measurements, the EO system was adjusted such that the phase step has the maximum possible effect on the signal. So, the bunch shape is not reproduced optimally. Instead, the frequency of the forced synchrotron oscillation after the RF phase step can be seen clearly. This enables to take the system into account for the detector synchronization. 

Generally, the EOSD system can also be set up to allow measurements of the bunch profile with a sub-ps resolution, that allows to resolve sub-structures on the longitudinal bunch profile \cite{Funkner2018}. By using KALYPSO this setup allows turn-by-turn measurements \cite[Fig. 3]{ROTA_IBIC_16}.

\section{Detector synchronization \label{sec:synchronization}}

A straight forward way to measure synchronously the different bunch parameters is to feed the signals from the different detectors into one common DAQ system, e.g. an oscilloscope. This is suitable if all devices are located close to each other at a storage ring with compact design (e.g. \cite{PhysRevSTAB.4.054401}). In our case, this is not directly possible for several reasons. The first point is the location of the detector systems at different positions around the storage ring. In addition, at these systems dedicated post-processing of the raw-data is required. The FGC and the EOSD setup delivers 2D images while the KAPTURE system provides -- in the multi-detector mode -- one data point per bunch, turn, and detector. The acquisition has to be aligned temporally to enable correlation studies between derived parameters like CSR intensity, energy spread and bunch length.

To provide a common origin for the time axis with single-turn resolution, we need to synchronize the acquisition. The timing system \cite{Hofmann_IPAC_10} -- based on one event generator (EVG) and several event receivers (EVR) -- is used to generate a synchronized measurement trigger. 

Furthermore, the intrinsic hardware delays have to be taken into account. KAPTURE and KALYPSO sample continuously, independent of the coupled detectors. The acquisition trigger starts the data storage into memory and therefore an instantaneous synchronization of the recording is achieved.

For the FGC this is different: Due to the measurement controls and the inertia of the rotating mirror, a certain advance time is needed before the first sample can be recorded. To protect the sensor from overexposure, the rotating mirror has a hold position in which the light is not on the sensor. Thus the light spot has to be driven first onto the sensor to be recorded. This also ensures the measurement is taken in the regime of linear mirror motion.
The synchronization schemes takes these delays into account and compensates for them by triggering the FGC the required time before other devices (see \cite[Fig. 2]{KehrerIPAC2016}). 

The test of the synchronization requires a signal with a signature on the longitudinal and horizontal plane of the beam that all detectors are sensitive to. We used the low-level RF (LLRF) system to trigger a step in the RF phase. This sudden change leads to the onset of a synchrotron oscillation around the new longitudinal equilibrium, which all different measurement stations are sensitive to. For the calibration of KAPTURE, we used the visible light APD as a reference. This convenient reference is insensitive to fluctuations on the bunch profile that can occur following a sudden RF phase-step, which can affect the CSR signal recorded by Schottky detectors.

\begin{figure}[b]
\includegraphics[width=\columnwidth]{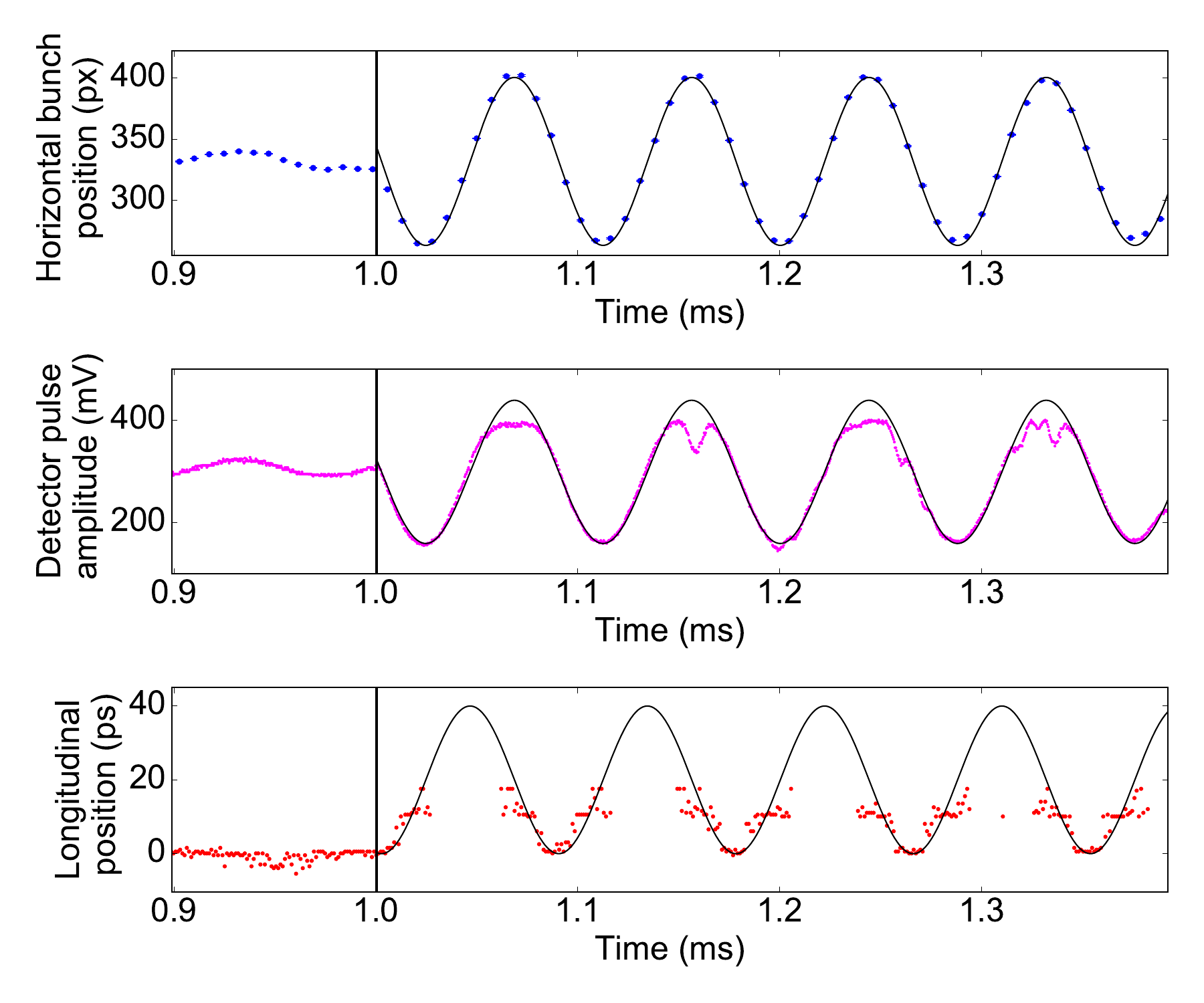}
\caption{Signature of the triggered RF phase-step seen by the different systems. Top: Horizontal bunch position from the FGC. Center: Signal of the incoherent synchrotron radiation recorded using an optical APD and KAPTURE. Bottom: Arrival time of the bunch recorded with KALYPSO. It is defined by the position of the maximum electro-optical modulation. The vertical line in all plots marks the RF phase step triggering the onset of a synchrotron oscillation. As a guide to the eye the plots contain a sinusoidal curve at the synchrotron frequency.}
\label{fig:Sync_3_detectors}
\end{figure}

Figure \ref{fig:Sync_3_detectors} shows a successful synchronization event. The onset of the synchrotron oscillation is recorded by all detectors at the same time. While the synchrotron oscillation on the FGC and the APD are in phase and both in the horizontal plane, the one from \mbox{KALYPSO} is phase-shifted by $\pi/2$. This can be explained by the dynamics in longitudinal phase space. Here, the synchrotron oscillation corresponds to a rotation of the longitudinal electron phase space density. The projection onto the position and energy axes are separated by $\pi/2$. The FGC and the APD are sensitive to the energy because their source points are located in dispersive sections. KALYPSO, in contrast, measures the arrival time.

The APD signal recorded with KAPTURE shows voltage dips at the highest values and maximum bunch deflection. The origin of the dips are unclear, but are likely due to intensity cut-off from the finite aperture of the optical beam path or jumps in the timing trigger as KAPTURE samples at a fixed phase. Nevertheless, the main feature, a strong synchrotron oscillation, can be seen and used to determine the phase and the frequency of the electron beam motion. 

As discussed before, the EOSD setup was optimized for temporal resolution. Due to the finite size of the line array and the length of the chirped laser pulse, this leads to a cut-off of the upper part of the signal showing the longitudinal bunch position.

\section{Synchronous measurements}

Synchronization of the detector systems allows time-resolved studies of the beam dynamics. In the following, we limit our studies to energy spread and the CSR emission during the micro-bunching instability. 

\begin{figure}[h]
\includegraphics[width=\columnwidth]{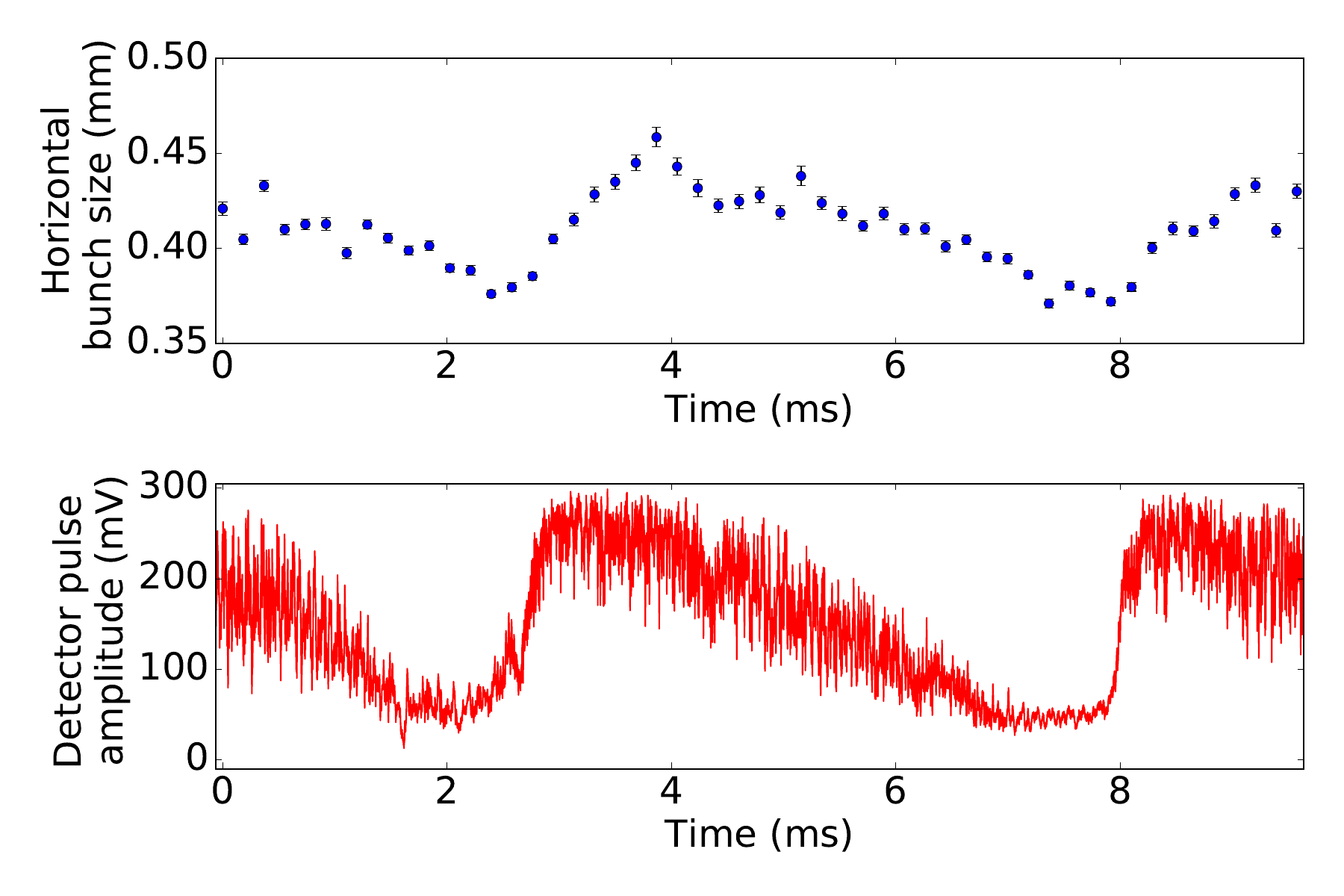}
\caption{Horizontal bunch size as measure for the energy spread (top) and the synchronously measured CSR signal sampled using a Schottky diode (bottom). The energy spread shows the same modulation pattern as the CSR (raw data published in \cite[Fig. 10]{Steinmann_FLS_2018}). \\ 
Beam parameter: $f_{\text{s}}$: \unit[8.1]{kHz}, $V_{\text{RF}}$: \unit[1500]{kV}, $I_{\text{bunch}}$: \unit[0.88]{mA}}
\label{fig:f05905_2016-04-08T01h02m15s_THzSignal_FGC_Crop_KAPTURE}
\end{figure}

\begin{figure}[h]
\includegraphics[width=\columnwidth]{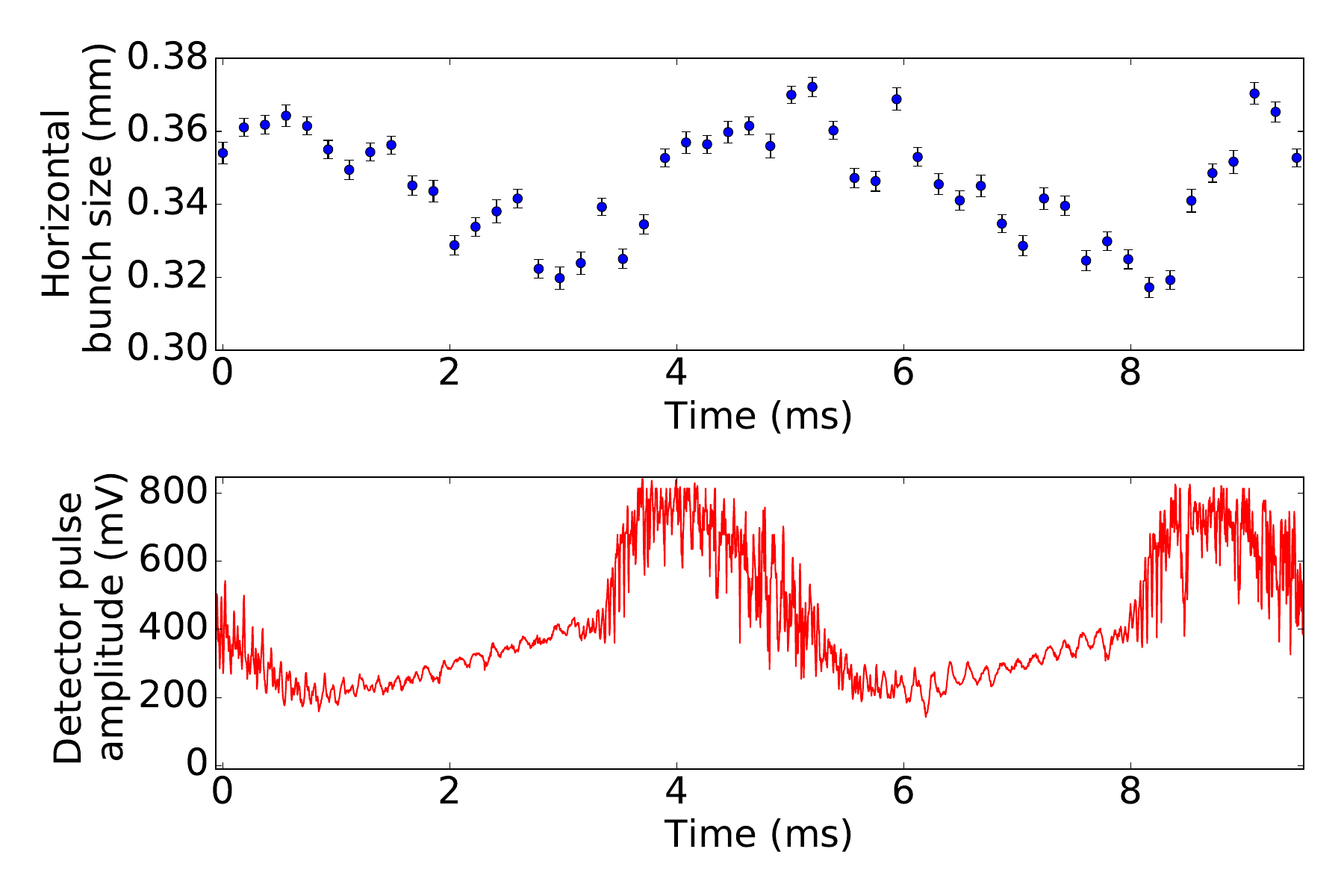}
\caption{Horizontal bunch size (top) and the synchronously measured CSR from a broadband Schottky diode (raw data published in \cite[Fig. 2]{kehrer2017time}).\\
Beam parameter: $f_{\text{s}}$: \unit[6.7]{kHz}, $V_{\text{RF}}$: \unit[1500]{kV}, $I_{\text{bunch}}$: \unit[0.25]{mA}.}
\label{fig:f06173_2017-01-19T01h57m31s_THzSignal_FGC_Crop_KAPTURE}
\end{figure}

Figures~\ref{fig:f05905_2016-04-08T01h02m15s_THzSignal_FGC_Crop_KAPTURE} and \ref{fig:f06173_2017-01-19T01h57m31s_THzSignal_FGC_Crop_KAPTURE} show two synchronous measurements of the horizontal bunch for the energy spread (top) and the corresponding CSR emission (bottom) recorded with a broadband Schottky diode detector (\unit[50]{GHz} - \unit[2]{THz}). In both cases, the two curves have the same modulation period. This behaviour was predicted by simulations \cite{Warnock:2006qa} and also observed experimentally at SURF III \cite{PhysRevSTAB.4.054401}.

In Fig.~\ref{fig:f05905_2016-04-08T01h02m15s_THzSignal_FGC_Crop_KAPTURE}, the CSR intensity decreases until it reaches a constant level between two bursts. After the blow-up at the onset of a burst, the energy spread decreases until it reaches the same lower limit where it is blown-up again. At the same time, the CSR burst starts to arise. 

Fig.~\ref{fig:f06173_2017-01-19T01h57m31s_THzSignal_FGC_Crop_KAPTURE} shows a different case using different machine settings and a lower bunch current. 
Here, the CSR intensity is not constant between two bursts, but increases slightly while the energy spread is still decreasing. This can be explained by the coupling of bunch length and energy spread in the longitudinal phase space. If the bunch gets shorter, it emits more CSR -- regardless of sub-structures.
The sensitivity to these bunch length fluctuations is dependent on the frequency band of the detector and possible shielding of the CSR due to apertures in the beam line. Thus, we do not observe it for longer bunches.

For both cases, the bunch contracts until an instability threshold is reached. At this point, the charge density inside the bunch is high enough and the bunch becomes unstable. Numerical simulations of the longitudinal phase space using the Vlasov-Fokker-Planck solver Inovesa \cite{PRAB_Inovesa} showed, that at this point the amplitude of the sub-structures starts to grow rapidly \cite{Boltz_master}. 

\begin{figure}[t]
\includegraphics[width=\columnwidth]{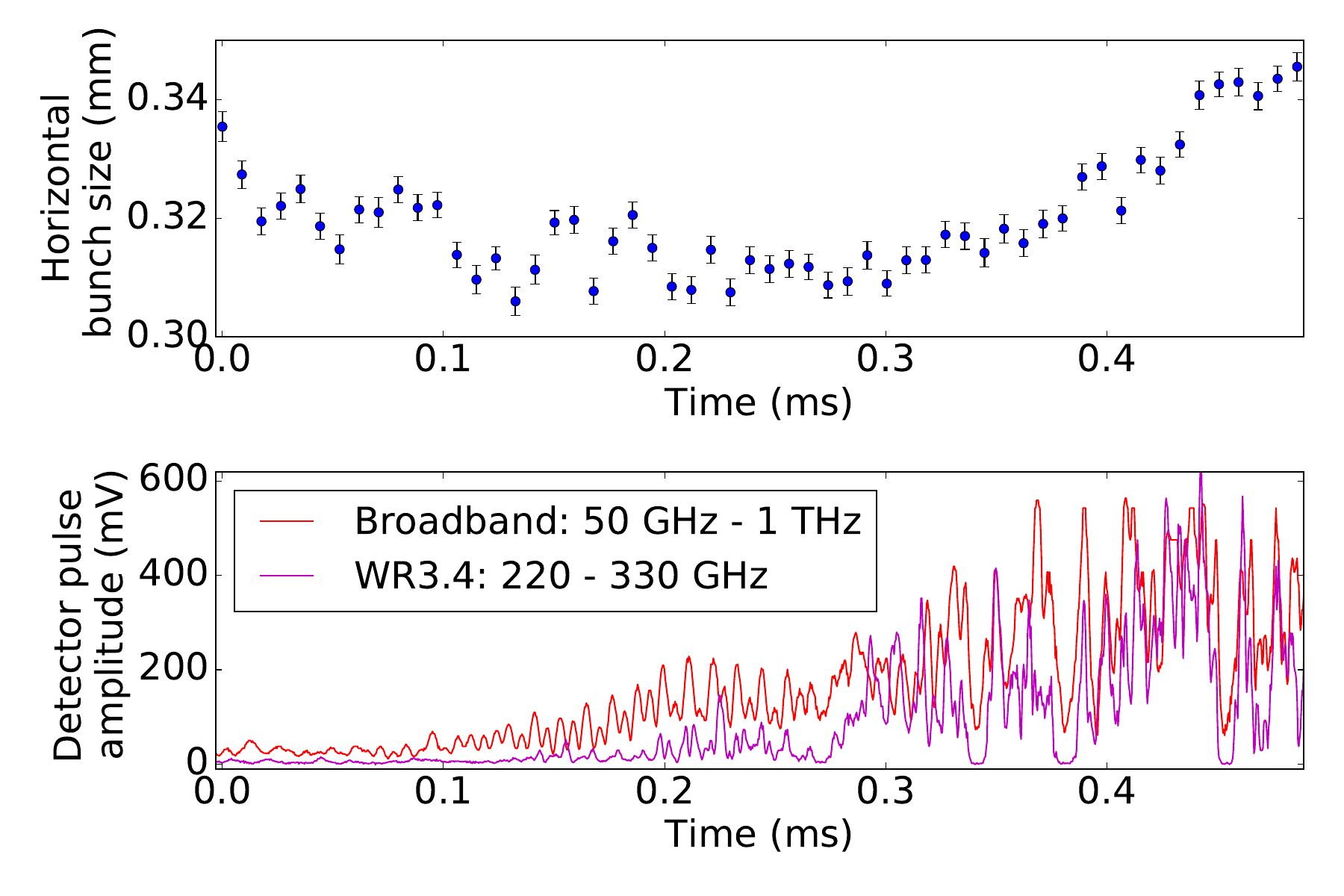}
\caption{Horizontal bunch size recorded with a gate separation of 24 turns (top) and the corresponding CSR intensity (bottom) sampled with two Schottky diodes with different bandwidths. The error bars for the spot size are large compared to the ones in Fig.~\ref{fig:f06173_2017-01-19T01h57m31s_THzSignal_FGC_Crop_KAPTURE} due to the lower amplification of the micro-channel plate leading to less counts on the CCD (raw data published in \cite[Fig. 3]{kehrer2017time}).\\
Beam parameter: $f_{\text{s}}$: \unit[13.4]{kHz}, $V_{\text{RF}}$: \unit[1500]{kV}, $I_{\text{bunch}}$: \unit[1.57]{mA}}
\label{fig:f05788_2015-11-06T03h29m46s_THzSignal_FGC_Crop_KAPTURE}
\end{figure}

To study these effects in more detail, the FGC can be configured to record shorter time ranges with a higher temporal resolution. For the third example discussed here, the gate separation was set to \unit[24]{turns} and a time range of \unit[500]{$\mu s$} was recorded. In this case an additional narrow-band Schottky diode was used synchronously.
The result is illustrated in Fig.~\ref{fig:f05788_2015-11-06T03h29m46s_THzSignal_FGC_Crop_KAPTURE}. At the beginning, the energy spread given by the horizontal bunch size (top panel) is slightly decreasing due to damping effects while the CSR intensity on the broadband Schottky diode is constant, for the narrow-band diode it is almost zero. 
Around t\ =\ \unit[0.1]{ms}, the first changes can be seen when the CSR intensity on the broadband Schottky diode starts to increase, while for the narrow-band this increase starts approximately \unit[0.1]{ms} later. The earlier increase of the signal on the broadband Schottky diode is due to its lower limit in the frequency sensitivity \cite{steinmann2017continuous}.

While the CSR intensity starts to increase, the energy spread is still decreasing due to damping effects. 
At \unit[0.27]{ms}, it reaches a lower limit triggering the onset of another burst. 

These examples clearly illustrate the diagnostic capability of synchronized detector systems. It allows to fully exploit the potential of the different, synchronous systems. While with KAPTURE in the multi-detector mode it is possible to study phase differences between the different frequency bands of the CSR \cite{steinmann2017continuous}, these phase studies can now be extended to take multiple detector systems into account. 

\section{Summary}
The analysis of fast rising instabilities, like the micro-bunching instability, benefits from turn-by-turn detector systems. To resolve complimentary features in 6-D phase space, several detector systems which observe different bunch parameters need to be synchronized.
At KARA, several independent detector systems with a single-turn resolution are combined in a distributed sensor network. The energy spread is investigated using an FGC, while the CSR intensity is measured using a KAPTURE system and Schottky diodes. For studies of the longitudinal bunch profile an EOSD setup is used in the near-field combined with KALYPSO. The integration of these systems into a hardware synchronization scheme opens up new diagnostics opportunities. 
While the individual setups allowed an insight into the dynamics of the micro-bunching instability, successful synchronization is a first step towards a multi-dimensional analysis of the longitudinal phase space. We experimentally showed that the energy spread and the CSR undergo a modulation with the same period length. In addition, detailed studies revealed a certain phase difference between the CSR in different frequency ranges and energy spread.

In the future we will improve and upgrade the individual detector systems while keeping the capability for synchronized recording. A fast line array optimized for the visible frequency range will be combined with KALYPSO and replace the FGC setup. This will allows us to continuously stream horizontal bunch profiles at each revolution. It will also be used at the EOSD setup with a line array, optimized for the laser frequency and improving the signal-to-noise ratio. Further, the next version of KALYPSO will support a streaming rate of up to \unit[10]{Mfps} and up to 2048 pixels \cite{CaselleIBIC2017}. A new version of KAPTURE, KAPTURE2 will provide 8 simultaneous readout paths, allowing more accurate peak reconstruction and the measurement of the arrival time \cite{Caselle2017KAPTURE}. Alternatively, the amplitudes of up to 8 different detectors can be recorded. With narrow-band detectors, sensitive in different frequency ranges, this allows recording single-shot spectra \cite{steinmann2017continuous}.

\section{Acknowledgements}

We would like to thank Anton Plech, Yves-Laurent Mathis and Michael Süpfle for their support. This work has been supported by the Initiative and Networking Fund of the Helmholtz Association under contract number VH-NG-320 and by the BMBF under contract numbers 05K10VKC and 05K13VKA.
Miriam Brosi, Patrik Schönfeldt and Johannes Steinmann acknowledge the support of the Helmholtz International Research School for Teratronics (HIRST) and Edmund Blomley the support of the Karlsruhe School of Elementary Particle and Astroparticle Physics (KSETA).

\bibliography{BibTex-reference/literature}

\begin{thebibliography}{41}%
\makeatletter
\providecommand \@ifxundefined [1]{%
 \@ifx{#1\undefined}
}%
\providecommand \@ifnum [1]{%
 \ifnum #1\expandafter \@firstoftwo
 \else \expandafter \@secondoftwo
 \fi
}%
\providecommand \@ifx [1]{%
 \ifx #1\expandafter \@firstoftwo
 \else \expandafter \@secondoftwo
 \fi
}%
\providecommand \natexlab [1]{#1}%
\providecommand \enquote  [1]{``#1''}%
\providecommand \bibnamefont  [1]{#1}%
\providecommand \bibfnamefont [1]{#1}%
\providecommand \citenamefont [1]{#1}%
\providecommand \href@noop [0]{\@secondoftwo}%
\providecommand \href [0]{\begingroup \@sanitize@url \@href}%
\providecommand \@href[1]{\@@startlink{#1}\@@href}%
\providecommand \@@href[1]{\endgroup#1\@@endlink}%
\providecommand \@sanitize@url [0]{\catcode `\\12\catcode `\$12\catcode
  `\&12\catcode `\#12\catcode `\^12\catcode `\_12\catcode `\%12\relax}%
\providecommand \@@startlink[1]{}%
\providecommand \@@endlink[0]{}%
\providecommand \url  [0]{\begingroup\@sanitize@url \@url }%
\providecommand \@url [1]{\endgroup\@href {#1}{\urlprefix }}%
\providecommand \urlprefix  [0]{URL }%
\providecommand \Eprint [0]{\href }%
\providecommand \doibase [0]{http://dx.doi.org/}%
\providecommand \selectlanguage [0]{\@gobble}%
\providecommand \bibinfo  [0]{\@secondoftwo}%
\providecommand \bibfield  [0]{\@secondoftwo}%
\providecommand \translation [1]{[#1]}%
\providecommand \BibitemOpen [0]{}%
\providecommand \bibitemStop [0]{}%
\providecommand \bibitemNoStop [0]{.\EOS\space}%
\providecommand \EOS [0]{\spacefactor3000\relax}%
\providecommand \BibitemShut  [1]{\csname bibitem#1\endcsname}%
\let\auto@bib@innerbib\@empty
%</preamble>
\bibitem [{\citenamefont {Carr}\ \emph {et~al.}(2001)\citenamefont {Carr},
  \citenamefont {Kramer}, \citenamefont {Murphy}, \citenamefont {Lobo},\ and\
  \citenamefont {Tanner}}]{CARR2001387}%
  \BibitemOpen
  \bibfield  {author} {\bibinfo {author} {\bibfnamefont {G.}~\bibnamefont
  {Carr}}, \bibinfo {author} {\bibfnamefont {S.}~\bibnamefont {Kramer}},
  \bibinfo {author} {\bibfnamefont {J.}~\bibnamefont {Murphy}}, \bibinfo
  {author} {\bibfnamefont {R.}~\bibnamefont {Lobo}}, \ and\ \bibinfo {author}
  {\bibfnamefont {D.}~\bibnamefont {Tanner}},\ }\href {\doibase
  10.1016/S0168-9002(01)00521-6} {\bibfield  {journal} {\bibinfo  {journal}
  {Nuclear Instruments and Methods in Physics Research Section A: Accelerators,
  Spectrometers, Detectors and Associated Equipment}\ }\textbf {\bibinfo
  {volume} {463}},\ \bibinfo {pages} {387 } (\bibinfo {year}
  {2001})}\BibitemShut {NoStop}%
\bibitem [{\citenamefont {Arp}\ \emph {et~al.}(2001)\citenamefont {Arp},
  \citenamefont {Fraser}, \citenamefont {Hight~Walker}, \citenamefont
  {Lucatorto}, \citenamefont {Lehmann}, \citenamefont {Harkay}, \citenamefont
  {Sereno},\ and\ \citenamefont {Kim}}]{PhysRevSTAB.4.054401}%
  \BibitemOpen
  \bibfield  {author} {\bibinfo {author} {\bibfnamefont {U.}~\bibnamefont
  {Arp}}, \bibinfo {author} {\bibfnamefont {G.~T.}\ \bibnamefont {Fraser}},
  \bibinfo {author} {\bibfnamefont {A.~R.}\ \bibnamefont {Hight~Walker}},
  \bibinfo {author} {\bibfnamefont {T.~B.}\ \bibnamefont {Lucatorto}}, \bibinfo
  {author} {\bibfnamefont {K.~K.}\ \bibnamefont {Lehmann}}, \bibinfo {author}
  {\bibfnamefont {K.}~\bibnamefont {Harkay}}, \bibinfo {author} {\bibfnamefont
  {N.}~\bibnamefont {Sereno}}, \ and\ \bibinfo {author} {\bibfnamefont {K.-J.}\
  \bibnamefont {Kim}},\ }\href {\doibase 10.1103/PhysRevSTAB.4.054401}
  {\bibfield  {journal} {\bibinfo  {journal} {Phys. Rev. ST Accel. Beams}\
  }\textbf {\bibinfo {volume} {4}},\ \bibinfo {pages} {054401} (\bibinfo {year}
  {2001})}\BibitemShut {NoStop}%
\bibitem [{\citenamefont {Byrd}\ \emph {et~al.}(2002)\citenamefont {Byrd},
  \citenamefont {Leemans}, \citenamefont {Loftsdottir}, \citenamefont
  {Marcelis}, \citenamefont {Martin}, \citenamefont {McKinney}, \citenamefont
  {Sannibale}, \citenamefont {Scarvie},\ and\ \citenamefont
  {Steier}}]{PhysRevLett.89.224801}%
  \BibitemOpen
  \bibfield  {author} {\bibinfo {author} {\bibfnamefont {J.~M.}\ \bibnamefont
  {Byrd}}, \bibinfo {author} {\bibfnamefont {W.~P.}\ \bibnamefont {Leemans}},
  \bibinfo {author} {\bibfnamefont {A.}~\bibnamefont {Loftsdottir}}, \bibinfo
  {author} {\bibfnamefont {B.}~\bibnamefont {Marcelis}}, \bibinfo {author}
  {\bibfnamefont {M.~C.}\ \bibnamefont {Martin}}, \bibinfo {author}
  {\bibfnamefont {W.~R.}\ \bibnamefont {McKinney}}, \bibinfo {author}
  {\bibfnamefont {F.}~\bibnamefont {Sannibale}}, \bibinfo {author}
  {\bibfnamefont {T.}~\bibnamefont {Scarvie}}, \ and\ \bibinfo {author}
  {\bibfnamefont {C.}~\bibnamefont {Steier}},\ }\href {\doibase
  10.1103/PhysRevLett.89.224801} {\bibfield  {journal} {\bibinfo  {journal}
  {Phys. Rev. Lett.}\ }\textbf {\bibinfo {volume} {89}},\ \bibinfo {pages}
  {224801} (\bibinfo {year} {2002})}\BibitemShut {NoStop}%
\bibitem [{\citenamefont {Abo-Bakr}\ \emph {et~al.}(2003)\citenamefont
  {Abo-Bakr}, \citenamefont {Feikes}, \citenamefont {Holldack}, \citenamefont
  {Kuske},\ and\ \citenamefont {W\"ustefeld}}]{abo2003coherent}%
  \BibitemOpen
  \bibfield  {author} {\bibinfo {author} {\bibfnamefont {M.}~\bibnamefont
  {Abo-Bakr}}, \bibinfo {author} {\bibfnamefont {J.}~\bibnamefont {Feikes}},
  \bibinfo {author} {\bibfnamefont {K.}~\bibnamefont {Holldack}}, \bibinfo
  {author} {\bibfnamefont {P.}~\bibnamefont {Kuske}}, \ and\ \bibinfo {author}
  {\bibfnamefont {G.}~\bibnamefont {W\"ustefeld}},\ }in\ \href {\doibase
  10.1109/PAC.2003.1289801} {\emph {\bibinfo {booktitle} {Proceedings of
  PAC'03}}},\ Vol.~\bibinfo {volume} {5}\ (\bibinfo {organization} {IEEE},\
  \bibinfo {year} {2003})\ pp.\ \bibinfo {pages} {3023--3025}\BibitemShut
  {NoStop}%
\bibitem [{\citenamefont {Shields}\ \emph {et~al.}(2012)\citenamefont
  {Shields}, \citenamefont {Bartolini}, \citenamefont {Boorman}, \citenamefont
  {Karataev}, \citenamefont {Lyapin}, \citenamefont {Puntree},\ and\
  \citenamefont {Rehm}}]{Shields_Diamond}%
  \BibitemOpen
  \bibfield  {author} {\bibinfo {author} {\bibfnamefont {W.}~\bibnamefont
  {Shields}}, \bibinfo {author} {\bibfnamefont {R.}~\bibnamefont {Bartolini}},
  \bibinfo {author} {\bibfnamefont {G.}~\bibnamefont {Boorman}}, \bibinfo
  {author} {\bibfnamefont {P.}~\bibnamefont {Karataev}}, \bibinfo {author}
  {\bibfnamefont {A.}~\bibnamefont {Lyapin}}, \bibinfo {author} {\bibfnamefont
  {J.}~\bibnamefont {Puntree}}, \ and\ \bibinfo {author} {\bibfnamefont
  {G.}~\bibnamefont {Rehm}},\ }\href {\doibase 10.1088/1742-6596/357/1/012037}
  {\bibfield  {journal} {\bibinfo  {journal} {Journal of Physics: Conference
  Series}\ }\textbf {\bibinfo {volume} {357}},\ \bibinfo {pages} {012037}
  (\bibinfo {year} {2012})}\BibitemShut {NoStop}%
\bibitem [{\citenamefont {Evain}\ \emph {et~al.}(2012)\citenamefont {Evain},
  \citenamefont {Barros}, \citenamefont {Loulergue}, \citenamefont {Tordeux},
  \citenamefont {Nagaoka}, \citenamefont {Labat}, \citenamefont {Cassinari},
  \citenamefont {Creff}, \citenamefont {Manceron}, \citenamefont {Brubach},
  \citenamefont {Roy},\ and\ \citenamefont {Couprie}}]{Evian_MBI_EPL}%
  \BibitemOpen
  \bibfield  {author} {\bibinfo {author} {\bibfnamefont {C.}~\bibnamefont
  {Evain}}, \bibinfo {author} {\bibfnamefont {J.}~\bibnamefont {Barros}},
  \bibinfo {author} {\bibfnamefont {A.}~\bibnamefont {Loulergue}}, \bibinfo
  {author} {\bibfnamefont {M.~A.}\ \bibnamefont {Tordeux}}, \bibinfo {author}
  {\bibfnamefont {R.}~\bibnamefont {Nagaoka}}, \bibinfo {author} {\bibfnamefont
  {M.}~\bibnamefont {Labat}}, \bibinfo {author} {\bibfnamefont
  {L.}~\bibnamefont {Cassinari}}, \bibinfo {author} {\bibfnamefont
  {G.}~\bibnamefont {Creff}}, \bibinfo {author} {\bibfnamefont
  {L.}~\bibnamefont {Manceron}}, \bibinfo {author} {\bibfnamefont {J.~B.}\
  \bibnamefont {Brubach}}, \bibinfo {author} {\bibfnamefont {P.}~\bibnamefont
  {Roy}}, \ and\ \bibinfo {author} {\bibfnamefont {M.~E.}\ \bibnamefont
  {Couprie}},\ }\href {\doibase 10.1209/0295-5075/98/40006} {\bibfield
  {journal} {\bibinfo  {journal} {EPL (Europhysics Letters)}\ }\textbf
  {\bibinfo {volume} {98}},\ \bibinfo {pages} {40006} (\bibinfo {year}
  {2012})}\BibitemShut {NoStop}%
\bibitem [{\citenamefont {Stupakov}\ and\ \citenamefont
  {Heifets}(2002)}]{PhysRevSTAB.5.054402}%
  \BibitemOpen
  \bibfield  {author} {\bibinfo {author} {\bibfnamefont {G.}~\bibnamefont
  {Stupakov}}\ and\ \bibinfo {author} {\bibfnamefont {S.}~\bibnamefont
  {Heifets}},\ }\href {\doibase 10.1103/PhysRevSTAB.5.054402} {\bibfield
  {journal} {\bibinfo  {journal} {Phys. Rev. ST Accel. Beams}\ }\textbf
  {\bibinfo {volume} {5}},\ \bibinfo {pages} {054402} (\bibinfo {year}
  {2002})}\BibitemShut {NoStop}%
\bibitem [{\citenamefont {Venturini}\ and\ \citenamefont
  {Warnock}(2002)}]{PhysRevLett.89.224802}%
  \BibitemOpen
  \bibfield  {author} {\bibinfo {author} {\bibfnamefont {M.}~\bibnamefont
  {Venturini}}\ and\ \bibinfo {author} {\bibfnamefont {R.}~\bibnamefont
  {Warnock}},\ }\href {\doibase 10.1103/PhysRevLett.89.224802} {\bibfield
  {journal} {\bibinfo  {journal} {Phys. Rev. Lett.}\ }\textbf {\bibinfo
  {volume} {89}},\ \bibinfo {pages} {224802} (\bibinfo {year}
  {2002})}\BibitemShut {NoStop}%
\bibitem [{\citenamefont {Warnock}(2006)}]{Warnock:2006qa}%
  \BibitemOpen
  \bibfield  {author} {\bibinfo {author} {\bibfnamefont {R.}~\bibnamefont
  {Warnock}},\ }\bibfield  {booktitle} {\emph {\bibinfo {booktitle} {{Workshop
  on High Intensity Beam Dynamics (Coulomb'05) Senigallia, Italy, September
  12-16, 2005}}},\ }\href {\doibase 10.1016/j.nima.2006.01.041} {\bibfield
  {journal} {\bibinfo  {journal} {Nucl. Instrum. Meth.}\ }\textbf {\bibinfo
  {volume} {A561}},\ \bibinfo {pages} {186} (\bibinfo {year}
  {2006})}\BibitemShut {NoStop}%
\bibitem [{\citenamefont {M{\"u}ller}\ \emph {et~al.}(2012)\citenamefont
  {M{\"u}ller}, \citenamefont {Hiller}, \citenamefont {A.}, \citenamefont
  {Huttel}, \citenamefont {Il'in}, \citenamefont {Judin}, \citenamefont
  {Kehrer}, \citenamefont {Klein}, \citenamefont {Marsching}, \citenamefont
  {Meuter}, \citenamefont {Naknaimueang}, \citenamefont {Nasse}, \citenamefont
  {Plech}, \citenamefont {Probst}, \citenamefont {Scheuring}, \citenamefont
  {Schuh}, \citenamefont {Schwarz}, \citenamefont {Siegel}, \citenamefont
  {Smale}, \citenamefont {Streichert}, \citenamefont {Caspers}, \citenamefont
  {Semenov}, \citenamefont {H{\"u}bers},\ and\ \citenamefont
  {Br{\"u}ndermann}}]{muller2012beam}%
  \BibitemOpen
  \bibfield  {author} {\bibinfo {author} {\bibfnamefont {A.-S.}\ \bibnamefont
  {M{\"u}ller}}, \bibinfo {author} {\bibfnamefont {N.}~\bibnamefont {Hiller}},
  \bibinfo {author} {\bibfnamefont {H.}~\bibnamefont {A.}}, \bibinfo {author}
  {\bibfnamefont {E.}~\bibnamefont {Huttel}}, \bibinfo {author} {\bibfnamefont
  {K.}~\bibnamefont {Il'in}}, \bibinfo {author} {\bibfnamefont
  {V.}~\bibnamefont {Judin}}, \bibinfo {author} {\bibfnamefont
  {B.}~\bibnamefont {Kehrer}}, \bibinfo {author} {\bibfnamefont
  {M.}~\bibnamefont {Klein}}, \bibinfo {author} {\bibfnamefont
  {S.}~\bibnamefont {Marsching}}, \bibinfo {author} {\bibfnamefont
  {C.}~\bibnamefont {Meuter}}, \bibinfo {author} {\bibfnamefont
  {S.}~\bibnamefont {Naknaimueang}}, \bibinfo {author} {\bibfnamefont
  {M.}~\bibnamefont {Nasse}}, \bibinfo {author} {\bibfnamefont
  {A.}~\bibnamefont {Plech}}, \bibinfo {author} {\bibfnamefont
  {P.}~\bibnamefont {Probst}}, \bibinfo {author} {\bibfnamefont
  {A.}~\bibnamefont {Scheuring}}, \bibinfo {author} {\bibfnamefont
  {M.}~\bibnamefont {Schuh}}, \bibinfo {author} {\bibfnamefont
  {M.}~\bibnamefont {Schwarz}}, \bibinfo {author} {\bibfnamefont
  {M.}~\bibnamefont {Siegel}}, \bibinfo {author} {\bibfnamefont
  {N.}~\bibnamefont {Smale}}, \bibinfo {author} {\bibfnamefont
  {M.}~\bibnamefont {Streichert}}, \bibinfo {author} {\bibfnamefont
  {F.}~\bibnamefont {Caspers}}, \bibinfo {author} {\bibfnamefont
  {A.}~\bibnamefont {Semenov}}, \bibinfo {author} {\bibfnamefont {H.-W.}\
  \bibnamefont {H{\"u}bers}}, \ and\ \bibinfo {author} {\bibfnamefont
  {E.}~\bibnamefont {Br{\"u}ndermann}},\ }\href@noop {} {\bibfield  {journal}
  {\bibinfo  {journal} {ICFA Beam dynamics newsletter}\ }\textbf {\bibinfo
  {volume} {57}},\ \bibinfo {pages} {154} (\bibinfo {year} {2012})}\BibitemShut
  {NoStop}%
\bibitem [{\citenamefont {Brosi}\ \emph {et~al.}(2016)\citenamefont {Brosi},
  \citenamefont {Steinmann}, \citenamefont {Blomley}, \citenamefont
  {Br{\"u}ndermann}, \citenamefont {Caselle}, \citenamefont {Hiller},
  \citenamefont {Kehrer}, \citenamefont {Mathis}, \citenamefont {Nasse},
  \citenamefont {Rota}, \citenamefont {Schedler}, \citenamefont
  {Sch{\"o}nfeldt}, \citenamefont {Schuh}, \citenamefont {Schwarz},
  \citenamefont {Weber},\ and\ \citenamefont
  {M{\"u}ller}}]{PhysRevAccelBeams.19.110701}%
  \BibitemOpen
  \bibfield  {author} {\bibinfo {author} {\bibfnamefont {M.}~\bibnamefont
  {Brosi}}, \bibinfo {author} {\bibfnamefont {J.~L.}\ \bibnamefont
  {Steinmann}}, \bibinfo {author} {\bibfnamefont {E.}~\bibnamefont {Blomley}},
  \bibinfo {author} {\bibfnamefont {E.}~\bibnamefont {Br{\"u}ndermann}},
  \bibinfo {author} {\bibfnamefont {M.}~\bibnamefont {Caselle}}, \bibinfo
  {author} {\bibfnamefont {N.}~\bibnamefont {Hiller}}, \bibinfo {author}
  {\bibfnamefont {B.}~\bibnamefont {Kehrer}}, \bibinfo {author} {\bibfnamefont
  {Y.-L.}\ \bibnamefont {Mathis}}, \bibinfo {author} {\bibfnamefont {M.~J.}\
  \bibnamefont {Nasse}}, \bibinfo {author} {\bibfnamefont {L.}~\bibnamefont
  {Rota}}, \bibinfo {author} {\bibfnamefont {M.}~\bibnamefont {Schedler}},
  \bibinfo {author} {\bibfnamefont {P.}~\bibnamefont {Sch{\"o}nfeldt}},
  \bibinfo {author} {\bibfnamefont {M.}~\bibnamefont {Schuh}}, \bibinfo
  {author} {\bibfnamefont {M.}~\bibnamefont {Schwarz}}, \bibinfo {author}
  {\bibfnamefont {M.}~\bibnamefont {Weber}}, \ and\ \bibinfo {author}
  {\bibfnamefont {A.-S.}\ \bibnamefont {M{\"u}ller}},\ }\href {\doibase
  10.1103/PhysRevAccelBeams.19.110701} {\bibfield  {journal} {\bibinfo
  {journal} {Phys. Rev. Accel. Beams}\ }\textbf {\bibinfo {volume} {19}},\
  \bibinfo {pages} {110701} (\bibinfo {year} {2016})}\BibitemShut {NoStop}%
\bibitem [{\citenamefont {Steinmann}\ \emph {et~al.}(2017)\citenamefont
  {Steinmann}, \citenamefont {Brosi}, \citenamefont {Br\"undermann},
  \citenamefont {Caselle}, \citenamefont {Kehrer}, \citenamefont {Rota},
  \citenamefont {Sch\"onfeldt}, \citenamefont {Schuh}, \citenamefont {Siegel},
  \citenamefont {Weber},\ and\ \citenamefont
  {M\"uller}}]{steinmann2017continuous}%
  \BibitemOpen
  \bibfield  {author} {\bibinfo {author} {\bibfnamefont {J.~L.}\ \bibnamefont
  {Steinmann}}, \bibinfo {author} {\bibfnamefont {M.}~\bibnamefont {Brosi}},
  \bibinfo {author} {\bibfnamefont {E.}~\bibnamefont {Br\"undermann}}, \bibinfo
  {author} {\bibfnamefont {M.}~\bibnamefont {Caselle}}, \bibinfo {author}
  {\bibfnamefont {B.}~\bibnamefont {Kehrer}}, \bibinfo {author} {\bibfnamefont
  {L.}~\bibnamefont {Rota}}, \bibinfo {author} {\bibfnamefont {P.}~\bibnamefont
  {Sch\"onfeldt}}, \bibinfo {author} {\bibfnamefont {M.}~\bibnamefont {Schuh}},
  \bibinfo {author} {\bibfnamefont {M.}~\bibnamefont {Siegel}}, \bibinfo
  {author} {\bibfnamefont {M.}~\bibnamefont {Weber}}, \ and\ \bibinfo {author}
  {\bibfnamefont {A.-S.}\ \bibnamefont {M\"uller}},\ }\href@noop {} {\bibfield
  {journal} {\bibinfo  {journal} {submitted to PRAB, preprint on arXiv}\ }
  (\bibinfo {year} {2017})},\ \Eprint {http://arxiv.org/abs/1710.09568}
  {arXiv:1710.09568} \BibitemShut {NoStop}%
\bibitem [{\citenamefont {Fisher}\ \emph {et~al.}(2006)\citenamefont {Fisher},
  \citenamefont {Petree}, \citenamefont {Kraus}, \citenamefont {Au},
  \citenamefont {Chan}, \citenamefont {Meyer},\ and\ \citenamefont
  {Webber}}]{fisher2006turn}%
  \BibitemOpen
  \bibfield  {author} {\bibinfo {author} {\bibfnamefont {A.}~\bibnamefont
  {Fisher}}, \bibinfo {author} {\bibfnamefont {M.}~\bibnamefont {Petree}},
  \bibinfo {author} {\bibfnamefont {R.}~\bibnamefont {Kraus}}, \bibinfo
  {author} {\bibfnamefont {Y.-S.}\ \bibnamefont {Au}}, \bibinfo {author}
  {\bibfnamefont {B.}~\bibnamefont {Chan}}, \bibinfo {author} {\bibfnamefont
  {T.}~\bibnamefont {Meyer}}, \ and\ \bibinfo {author} {\bibfnamefont
  {R.}~\bibnamefont {Webber}},\ }in\ \href {\doibase 10.1063/1.2401418} {\emph
  {\bibinfo {booktitle} {AIP Conference Proceedings}}},\ Vol.\ \bibinfo
  {volume} {868}\ (\bibinfo {organization} {AIP},\ \bibinfo {year} {2006})\
  pp.\ \bibinfo {pages} {303--312}\BibitemShut {NoStop}%
\bibitem [{\citenamefont {H{\"u}bers}\ \emph {et~al.}(2005)\citenamefont
  {H{\"u}bers}, \citenamefont {Semenov}, \citenamefont {Holldack},
  \citenamefont {Schade}, \citenamefont {W{\"u}stefeld},\ and\ \citenamefont
  {Gol’tsman}}]{hubers2005time}%
  \BibitemOpen
  \bibfield  {author} {\bibinfo {author} {\bibfnamefont {H.-W.}\ \bibnamefont
  {H{\"u}bers}}, \bibinfo {author} {\bibfnamefont {A.}~\bibnamefont {Semenov}},
  \bibinfo {author} {\bibfnamefont {K.}~\bibnamefont {Holldack}}, \bibinfo
  {author} {\bibfnamefont {U.}~\bibnamefont {Schade}}, \bibinfo {author}
  {\bibfnamefont {G.}~\bibnamefont {W{\"u}stefeld}}, \ and\ \bibinfo {author}
  {\bibfnamefont {G.}~\bibnamefont {Gol’tsman}},\ }\href {\doibase
  10.1063/1.2120896} {\bibfield  {journal} {\bibinfo  {journal} {Applied
  Physics Letters}\ }\textbf {\bibinfo {volume} {87}},\ \bibinfo {pages}
  {184103} (\bibinfo {year} {2005})}\BibitemShut {NoStop}%
\bibitem [{\citenamefont {Wilke}\ \emph {et~al.}(2002)\citenamefont {Wilke},
  \citenamefont {MacLeod}, \citenamefont {Gillespie}, \citenamefont {Berden},
  \citenamefont {Knippels},\ and\ \citenamefont {van~der
  Meer}}]{Wilke_EOSD_SingleShot_PRL}%
  \BibitemOpen
  \bibfield  {author} {\bibinfo {author} {\bibfnamefont {I.}~\bibnamefont
  {Wilke}}, \bibinfo {author} {\bibfnamefont {A.~M.}\ \bibnamefont {MacLeod}},
  \bibinfo {author} {\bibfnamefont {W.~A.}\ \bibnamefont {Gillespie}}, \bibinfo
  {author} {\bibfnamefont {G.}~\bibnamefont {Berden}}, \bibinfo {author}
  {\bibfnamefont {G.~M.~H.}\ \bibnamefont {Knippels}}, \ and\ \bibinfo {author}
  {\bibfnamefont {A.~F.~G.}\ \bibnamefont {van~der Meer}},\ }\href {\doibase
  10.1103/PhysRevLett.88.124801} {\bibfield  {journal} {\bibinfo  {journal}
  {Phys. Rev. Lett.}\ }\textbf {\bibinfo {volume} {88}},\ \bibinfo {pages}
  {124801} (\bibinfo {year} {2002})}\BibitemShut {NoStop}%
\bibitem [{\citenamefont {Wiedemann}(2007)}]{wiedemann2007particle}%
  \BibitemOpen
  \bibfield  {author} {\bibinfo {author} {\bibfnamefont {H.}~\bibnamefont
  {Wiedemann}},\ }\href {\doibase 10.1007/978-3-540-49045-6} {\emph {\bibinfo
  {title} {{Particle accelerator physics}}}}\ (\bibinfo  {publisher} {Springer
  Science \& Business Media},\ \bibinfo {year} {2007})\BibitemShut {NoStop}%
\bibitem [{\citenamefont {Kehrer}\ \emph {et~al.}(2015)\citenamefont {Kehrer},
  \citenamefont {Borysenko}, \citenamefont {Hertle}, \citenamefont {Hiller},
  \citenamefont {Holz}, \citenamefont {M{\"u}ller}, \citenamefont
  {Sch{\"o}nfeldt},\ and\ \citenamefont {Sch{\"u}tze}}]{KehrerIPAC2015}%
  \BibitemOpen
  \bibfield  {author} {\bibinfo {author} {\bibfnamefont {B.}~\bibnamefont
  {Kehrer}}, \bibinfo {author} {\bibfnamefont {A.}~\bibnamefont {Borysenko}},
  \bibinfo {author} {\bibfnamefont {E.}~\bibnamefont {Hertle}}, \bibinfo
  {author} {\bibfnamefont {N.}~\bibnamefont {Hiller}}, \bibinfo {author}
  {\bibfnamefont {M.}~\bibnamefont {Holz}}, \bibinfo {author} {\bibfnamefont
  {A.-S.}\ \bibnamefont {M{\"u}ller}}, \bibinfo {author} {\bibfnamefont
  {P.}~\bibnamefont {Sch{\"o}nfeldt}}, \ and\ \bibinfo {author} {\bibfnamefont
  {P.}~\bibnamefont {Sch{\"u}tze}},\ }in\ \href {\doibase
  10.18429/JACoW-IPAC2015-MOPHA037} {\emph {\bibinfo {booktitle} {Proceedings
  of IPAC'15}}}\ (\bibinfo {year} {2015})\ pp.\ \bibinfo {pages}
  {866--868}\BibitemShut {NoStop}%
\bibitem [{\citenamefont {Sch{\"u}tze}\ \emph {et~al.}(2015)\citenamefont
  {Sch{\"u}tze}, \citenamefont {Borysenko}, \citenamefont {Hertle},
  \citenamefont {Hiller}, \citenamefont {Kehrer}, \citenamefont {M{\"u}ller},\
  and\ \citenamefont {Sch{\"o}nfeldt}}]{SchuetzeIPAC2015}%
  \BibitemOpen
  \bibfield  {author} {\bibinfo {author} {\bibfnamefont {P.}~\bibnamefont
  {Sch{\"u}tze}}, \bibinfo {author} {\bibfnamefont {A.}~\bibnamefont
  {Borysenko}}, \bibinfo {author} {\bibfnamefont {E.}~\bibnamefont {Hertle}},
  \bibinfo {author} {\bibfnamefont {N.}~\bibnamefont {Hiller}}, \bibinfo
  {author} {\bibfnamefont {B.}~\bibnamefont {Kehrer}}, \bibinfo {author}
  {\bibfnamefont {A.-S.}\ \bibnamefont {M{\"u}ller}}, \ and\ \bibinfo {author}
  {\bibfnamefont {P.}~\bibnamefont {Sch{\"o}nfeldt}},\ }in\ \href {\doibase
  10.18429/JACoW-IPAC2015-MOPHA039} {\emph {\bibinfo {booktitle} {Proceedings
  of IPAC'15}}}\ (\bibinfo {year} {2015})\ pp.\ \bibinfo {pages}
  {872--875}\BibitemShut {NoStop}%
\bibitem [{\citenamefont {Sch{\"u}tze}(2017)}]{Schuetze_master}%
  \BibitemOpen
  \bibfield  {author} {\bibinfo {author} {\bibfnamefont {P.}~\bibnamefont
  {Sch{\"u}tze}},\ }\href@noop {} {\emph {\bibinfo {title} {{Transversale
  Strahldynamik bei der Erzeugung koh{\"a}renter Synchrotronstrahlung}}}},\
  BestMasters\ (\bibinfo  {publisher} {Springer},\ \bibinfo {year}
  {2017})\BibitemShut {NoStop}%
\bibitem [{\citenamefont {Hiller}\ \emph {et~al.}(2011)\citenamefont {Hiller},
  \citenamefont {Hofmann}, \citenamefont {Huttel}, \citenamefont {Judin},
  \citenamefont {Kehrer}, \citenamefont {Klein}, \citenamefont {Marsching},\
  and\ \citenamefont {M{\"u}ller}}]{HillerIPAC2011}%
  \BibitemOpen
  \bibfield  {author} {\bibinfo {author} {\bibfnamefont {N.}~\bibnamefont
  {Hiller}}, \bibinfo {author} {\bibfnamefont {A.}~\bibnamefont {Hofmann}},
  \bibinfo {author} {\bibfnamefont {E.}~\bibnamefont {Huttel}}, \bibinfo
  {author} {\bibfnamefont {V.}~\bibnamefont {Judin}}, \bibinfo {author}
  {\bibfnamefont {B.}~\bibnamefont {Kehrer}}, \bibinfo {author} {\bibfnamefont
  {M.}~\bibnamefont {Klein}}, \bibinfo {author} {\bibfnamefont
  {S.}~\bibnamefont {Marsching}}, \ and\ \bibinfo {author} {\bibfnamefont
  {A.-S.}\ \bibnamefont {M{\"u}ller}},\ }in\ \href@noop {} {\emph {\bibinfo
  {booktitle} {Proceedings of IPAC'11}}}\ (\bibinfo {year} {2011})\ pp.\
  \bibinfo {pages} {2951--2953}\BibitemShut {NoStop}%
\bibitem [{\citenamefont {Hofmann}\ and\ \citenamefont
  {M{\'e}ot}(1982)}]{hofmann1982optical}%
  \BibitemOpen
  \bibfield  {author} {\bibinfo {author} {\bibfnamefont {A.}~\bibnamefont
  {Hofmann}}\ and\ \bibinfo {author} {\bibfnamefont {F.}~\bibnamefont
  {M{\'e}ot}},\ }\href {\doibase 10.1016/0167-5087(82)90663-9} {\bibfield
  {journal} {\bibinfo  {journal} {Nuclear Instruments and Methods in Physics
  Research}\ }\textbf {\bibinfo {volume} {203}},\ \bibinfo {pages} {483}
  (\bibinfo {year} {1982})}\BibitemShut {NoStop}%
\bibitem [{\citenamefont {Andersson}\ \emph {et~al.}(2006)\citenamefont
  {Andersson}, \citenamefont {Schlott}, \citenamefont {Rohrer}, \citenamefont
  {Streun},\ and\ \citenamefont {Chubar}}]{oleg2006electron}%
  \BibitemOpen
  \bibfield  {author} {\bibinfo {author} {\bibfnamefont {{\AA}.}~\bibnamefont
  {Andersson}}, \bibinfo {author} {\bibfnamefont {V.}~\bibnamefont {Schlott}},
  \bibinfo {author} {\bibfnamefont {M.}~\bibnamefont {Rohrer}}, \bibinfo
  {author} {\bibfnamefont {A.}~\bibnamefont {Streun}}, \ and\ \bibinfo {author}
  {\bibfnamefont {O.}~\bibnamefont {Chubar}},\ }in\ \href@noop {} {\emph
  {\bibinfo {booktitle} {Proceedings of EPAC'06}}}\ (\bibinfo {year} {2006})\
  pp.\ \bibinfo {pages} {1223 -- 1225}\BibitemShut {NoStop}%
\bibitem [{Note1()}]{Note1}%
  \BibitemOpen
  \bibinfo {note} {OpTalix Pro 8.82, www.optenso.com}\BibitemShut {NoStop}%
\bibitem [{\citenamefont {Bane}\ \emph {et~al.}(2005)\citenamefont {Bane},
  \citenamefont {Oide},\ and\ \citenamefont {Zobov}}]{Bane2005}%
  \BibitemOpen
  \bibfield  {author} {\bibinfo {author} {\bibfnamefont {K.~L.~F.}\
  \bibnamefont {Bane}}, \bibinfo {author} {\bibfnamefont {K.}~\bibnamefont
  {Oide}}, \ and\ \bibinfo {author} {\bibfnamefont {M.}~\bibnamefont {Zobov}},\
  }in\ \href {\doibase 10.5170/CERN-2005-006.143} {\emph {\bibinfo {booktitle}
  {1st CARE-HHH-APD Workshop on Beam Dynamics in Future Hadron Colliders and
  Rapidly Cycling High-Intensity Synchrotrons}}}\ (\bibinfo {year}
  {2005})\BibitemShut {NoStop}%
\bibitem [{\citenamefont {Nash}\ \emph {et~al.}(2015)\citenamefont {Nash},
  \citenamefont {Carmignani}, \citenamefont {Farvacque}, \citenamefont
  {Liuzzo}, \citenamefont {Perron}, \citenamefont {Raimondi}, \citenamefont
  {Versteegen},\ and\ \citenamefont {White}}]{nash2015new}%
  \BibitemOpen
  \bibfield  {author} {\bibinfo {author} {\bibfnamefont {B.}~\bibnamefont
  {Nash}}, \bibinfo {author} {\bibfnamefont {N.}~\bibnamefont {Carmignani}},
  \bibinfo {author} {\bibfnamefont {L.}~\bibnamefont {Farvacque}}, \bibinfo
  {author} {\bibfnamefont {S.}~\bibnamefont {Liuzzo}}, \bibinfo {author}
  {\bibfnamefont {T.}~\bibnamefont {Perron}}, \bibinfo {author} {\bibfnamefont
  {P.}~\bibnamefont {Raimondi}}, \bibinfo {author} {\bibfnamefont
  {R.}~\bibnamefont {Versteegen}}, \ and\ \bibinfo {author} {\bibfnamefont
  {S.}~\bibnamefont {White}},\ }in\ \href@noop {} {\emph {\bibinfo {booktitle}
  {Proceedings of IPAC'15}}}\ (\bibinfo {year} {2015})\ pp.\ \bibinfo {pages}
  {113--116}\BibitemShut {NoStop}%
\bibitem [{\citenamefont {Safranek}\ \emph {et~al.}(2009)\citenamefont
  {Safranek}, \citenamefont {Portmann},\ and\ \citenamefont
  {Huang}}]{safranek2009linear}%
  \BibitemOpen
  \bibfield  {author} {\bibinfo {author} {\bibfnamefont {J.}~\bibnamefont
  {Safranek}}, \bibinfo {author} {\bibfnamefont {G.}~\bibnamefont {Portmann}},
  \ and\ \bibinfo {author} {\bibfnamefont {X.}~\bibnamefont {Huang}},\
  }\href@noop {} {\bibfield  {journal} {\bibinfo  {journal} {ICFA Beam Dynamics
  Newsletter}\ }\textbf {\bibinfo {volume} {44}},\ \bibinfo {pages} {43 }
  (\bibinfo {year} {2009})}\BibitemShut {NoStop}%
\bibitem [{\citenamefont {Caselle}\ \emph {et~al.}(2014)\citenamefont
  {Caselle}, \citenamefont {Brosi}, \citenamefont {Chilingaryan}, \citenamefont
  {Dritschler}, \citenamefont {Judin}, \citenamefont {Kopmann}, \citenamefont
  {M{\"u}ller}, \citenamefont {Raasch}, \citenamefont {Smale}, \citenamefont
  {Steinmann}, \citenamefont {Vogelgesang}, \citenamefont {Wuensch},
  \citenamefont {Siegel},\ and\ \citenamefont {Weber}}]{caselle2014ultra}%
  \BibitemOpen
  \bibfield  {author} {\bibinfo {author} {\bibfnamefont {M.}~\bibnamefont
  {Caselle}}, \bibinfo {author} {\bibfnamefont {M.}~\bibnamefont {Brosi}},
  \bibinfo {author} {\bibfnamefont {S.}~\bibnamefont {Chilingaryan}}, \bibinfo
  {author} {\bibfnamefont {T.}~\bibnamefont {Dritschler}}, \bibinfo {author}
  {\bibfnamefont {V.}~\bibnamefont {Judin}}, \bibinfo {author} {\bibfnamefont
  {A.}~\bibnamefont {Kopmann}}, \bibinfo {author} {\bibfnamefont {A.-S.}\
  \bibnamefont {M{\"u}ller}}, \bibinfo {author} {\bibfnamefont
  {J.}~\bibnamefont {Raasch}}, \bibinfo {author} {\bibfnamefont
  {N.}~\bibnamefont {Smale}}, \bibinfo {author} {\bibfnamefont
  {J.}~\bibnamefont {Steinmann}}, \bibinfo {author} {\bibfnamefont
  {M.}~\bibnamefont {Vogelgesang}}, \bibinfo {author} {\bibfnamefont
  {S.}~\bibnamefont {Wuensch}}, \bibinfo {author} {\bibfnamefont
  {M.}~\bibnamefont {Siegel}}, \ and\ \bibinfo {author} {\bibfnamefont
  {M.}~\bibnamefont {Weber}},\ }in\ \href {\doibase 10.1109/RTC.2014.7097535}
  {\emph {\bibinfo {booktitle} {Real Time Conference (RT), 2014 19th
  IEEE-NPSS}}}\ (\bibinfo {organization} {IEEE},\ \bibinfo {year} {2014})\ pp.\
  \bibinfo {pages} {1--3}\BibitemShut {NoStop}%
\bibitem [{\citenamefont {Caselle}\ \emph
  {et~al.}(2017{\natexlab{a}})\citenamefont {Caselle}, \citenamefont
  {Ardila~Perez}, \citenamefont {Balzer}, \citenamefont {Kopmann},
  \citenamefont {Rota}, \citenamefont {Weber}, \citenamefont {Brosi},
  \citenamefont {Steinmann}, \citenamefont {Br{\"u}ndermann},\ and\
  \citenamefont {M{\"u}ller}}]{Caselle2017KAPTURE}%
  \BibitemOpen
  \bibfield  {author} {\bibinfo {author} {\bibfnamefont {M.}~\bibnamefont
  {Caselle}}, \bibinfo {author} {\bibfnamefont {L.}~\bibnamefont
  {Ardila~Perez}}, \bibinfo {author} {\bibfnamefont {M.}~\bibnamefont
  {Balzer}}, \bibinfo {author} {\bibfnamefont {A.}~\bibnamefont {Kopmann}},
  \bibinfo {author} {\bibfnamefont {L.}~\bibnamefont {Rota}}, \bibinfo {author}
  {\bibfnamefont {M.}~\bibnamefont {Weber}}, \bibinfo {author} {\bibfnamefont
  {M.}~\bibnamefont {Brosi}}, \bibinfo {author} {\bibfnamefont
  {J.}~\bibnamefont {Steinmann}}, \bibinfo {author} {\bibfnamefont
  {E.}~\bibnamefont {Br{\"u}ndermann}}, \ and\ \bibinfo {author} {\bibfnamefont
  {A.-S.}\ \bibnamefont {M{\"u}ller}},\ }\href@noop {} {\bibfield  {journal}
  {\bibinfo  {journal} {Journal of Instrumentation}\ }\textbf {\bibinfo
  {volume} {12}},\ \bibinfo {pages} {C01040} (\bibinfo {year}
  {2017}{\natexlab{a}})}\BibitemShut {NoStop}%
\bibitem [{\citenamefont {{Virginia Diodes, Inc.}}()}]{VDI}%
  \BibitemOpen
  \bibfield  {author} {\bibinfo {author} {\bibnamefont {{Virginia Diodes,
  Inc.}}},\ }\href@noop {} {}\bibinfo {howpublished}
  {\url{http://vadiodes.com/index.php/en/}}\BibitemShut {NoStop}%
\bibitem [{\citenamefont {Jiang}\ and\ \citenamefont
  {Zhang}(1998)}]{ZhangFirstEOSD}%
  \BibitemOpen
  \bibfield  {author} {\bibinfo {author} {\bibfnamefont {Z.}~\bibnamefont
  {Jiang}}\ and\ \bibinfo {author} {\bibfnamefont {X.-C.}\ \bibnamefont
  {Zhang}},\ }\href {\doibase 10.1063/1.121231} {\bibfield  {journal} {\bibinfo
   {journal} {Applied Physics Letters}\ }\textbf {\bibinfo {volume} {72}},\
  \bibinfo {pages} {1945} (\bibinfo {year} {1998})}\BibitemShut {NoStop}%
\bibitem [{\citenamefont {Roussel}\ \emph {et~al.}(2016)\citenamefont
  {Roussel}, \citenamefont {Bielawski}, \citenamefont {Borysenko},
  \citenamefont {Evain}, \citenamefont {Hiller}, \citenamefont {M{\"u}ller},
  \citenamefont {Sch{\"o}nfeldt}, \citenamefont {Steinmann},\ and\
  \citenamefont {Szwaj}}]{roussel2016electro}%
  \BibitemOpen
  \bibfield  {author} {\bibinfo {author} {\bibfnamefont {E.}~\bibnamefont
  {Roussel}}, \bibinfo {author} {\bibfnamefont {S.}~\bibnamefont {Bielawski}},
  \bibinfo {author} {\bibfnamefont {A.}~\bibnamefont {Borysenko}}, \bibinfo
  {author} {\bibfnamefont {C.}~\bibnamefont {Evain}}, \bibinfo {author}
  {\bibfnamefont {N.}~\bibnamefont {Hiller}}, \bibinfo {author} {\bibfnamefont
  {A.-S.}\ \bibnamefont {M{\"u}ller}}, \bibinfo {author} {\bibfnamefont
  {P.}~\bibnamefont {Sch{\"o}nfeldt}}, \bibinfo {author} {\bibfnamefont
  {J.~L.}\ \bibnamefont {Steinmann}}, \ and\ \bibinfo {author} {\bibfnamefont
  {C.}~\bibnamefont {Szwaj}},\ }in\ \href@noop {} {\emph {\bibinfo {booktitle}
  {Proceedings of IBIC'15}}}\ (\bibinfo {year} {2016})\ pp.\ \bibinfo {pages}
  {33--37}\BibitemShut {NoStop}%
\bibitem [{\citenamefont {Hiller}\ \emph {et~al.}(2014)\citenamefont {Hiller},
  \citenamefont {Borysenko}, \citenamefont {Hertle}, \citenamefont {Judin},
  \citenamefont {Kehrer}, \citenamefont {Marsching}, \citenamefont
  {M{\"u}ller},\ and\ \citenamefont {Nasse}}]{hillerIPAC2014}%
  \BibitemOpen
  \bibfield  {author} {\bibinfo {author} {\bibfnamefont {N.}~\bibnamefont
  {Hiller}}, \bibinfo {author} {\bibfnamefont {A.}~\bibnamefont {Borysenko}},
  \bibinfo {author} {\bibfnamefont {E.}~\bibnamefont {Hertle}}, \bibinfo
  {author} {\bibfnamefont {V.}~\bibnamefont {Judin}}, \bibinfo {author}
  {\bibfnamefont {B.}~\bibnamefont {Kehrer}}, \bibinfo {author} {\bibfnamefont
  {S.}~\bibnamefont {Marsching}}, \bibinfo {author} {\bibfnamefont
  {A.}~\bibnamefont {M{\"u}ller}}, \ and\ \bibinfo {author} {\bibfnamefont
  {M.}~\bibnamefont {Nasse}},\ }in\ \href@noop {} {\emph {\bibinfo {booktitle}
  {Proceedings of IPAC'14}}}\ (\bibinfo {year} {2014})\BibitemShut {NoStop}%
\bibitem [{\citenamefont {Rota}\ \emph {et~al.}(2016)\citenamefont {Rota},
  \citenamefont {Balzer}, \citenamefont {Caselle}, \citenamefont {Weber},
  \citenamefont {Niehues}, \citenamefont {Sch{\"o}nfeldt}, \citenamefont
  {Nasse}, \citenamefont {M{\"u}ller}, \citenamefont {Hiller}, \citenamefont
  {Mozzanica}, \citenamefont {Gerth}, \citenamefont {Steffen}, \citenamefont
  {Makowski},\ and\ \citenamefont {Mielczarek}}]{ROTA_IBIC_16}%
  \BibitemOpen
  \bibfield  {author} {\bibinfo {author} {\bibfnamefont {L.}~\bibnamefont
  {Rota}}, \bibinfo {author} {\bibfnamefont {M.}~\bibnamefont {Balzer}},
  \bibinfo {author} {\bibfnamefont {M.}~\bibnamefont {Caselle}}, \bibinfo
  {author} {\bibfnamefont {M.}~\bibnamefont {Weber}}, \bibinfo {author}
  {\bibfnamefont {G.}~\bibnamefont {Niehues}}, \bibinfo {author} {\bibfnamefont
  {P.}~\bibnamefont {Sch{\"o}nfeldt}}, \bibinfo {author} {\bibfnamefont
  {M.}~\bibnamefont {Nasse}}, \bibinfo {author} {\bibfnamefont {A.-S.}\
  \bibnamefont {M{\"u}ller}}, \bibinfo {author} {\bibfnamefont
  {N.}~\bibnamefont {Hiller}}, \bibinfo {author} {\bibfnamefont
  {A.}~\bibnamefont {Mozzanica}}, \bibinfo {author} {\bibfnamefont
  {C.}~\bibnamefont {Gerth}}, \bibinfo {author} {\bibfnamefont
  {B.}~\bibnamefont {Steffen}}, \bibinfo {author} {\bibfnamefont
  {D.}~\bibnamefont {Makowski}}, \ and\ \bibinfo {author} {\bibfnamefont
  {A.}~\bibnamefont {Mielczarek}},\ }in\ \href {\doibase
  10.18429/JACoW-IBIC2016-WEPG46} {\emph {\bibinfo {booktitle} {Proceedings of
  IBIC'16}}}\ (\bibinfo {year} {2016})\ pp.\ \bibinfo {pages} {740 --
  743}\BibitemShut {NoStop}%
\bibitem [{\citenamefont {Caselle}\ \emph
  {et~al.}(2017{\natexlab{b}})\citenamefont {Caselle}, \citenamefont {Rota},
  \citenamefont {Balzer}, \citenamefont {Brosi}, \citenamefont {Funkner},
  \citenamefont {Kehrer}, \citenamefont {Nasse}, \citenamefont {Niehuse},
  \citenamefont {Patil}, \citenamefont {Sch{\"o}nfeldt}, \citenamefont {Schuh},
  \citenamefont {Steinmann}, \citenamefont {Yan}, \citenamefont
  {Br{\"u}ndermann}, \citenamefont {Weber}, \citenamefont {M{\"u}ller},
  \citenamefont {Borghi}, \citenamefont {Boscardin},\ and\ \citenamefont
  {Ronchin}}]{CaselleIBIC2017}%
  \BibitemOpen
  \bibfield  {author} {\bibinfo {author} {\bibfnamefont {M.}~\bibnamefont
  {Caselle}}, \bibinfo {author} {\bibfnamefont {L.}~\bibnamefont {Rota}},
  \bibinfo {author} {\bibfnamefont {M.}~\bibnamefont {Balzer}}, \bibinfo
  {author} {\bibfnamefont {M.}~\bibnamefont {Brosi}}, \bibinfo {author}
  {\bibfnamefont {S.}~\bibnamefont {Funkner}}, \bibinfo {author} {\bibfnamefont
  {B.}~\bibnamefont {Kehrer}}, \bibinfo {author} {\bibfnamefont
  {M.}~\bibnamefont {Nasse}}, \bibinfo {author} {\bibfnamefont
  {G.}~\bibnamefont {Niehuse}}, \bibinfo {author} {\bibfnamefont
  {M.}~\bibnamefont {Patil}}, \bibinfo {author} {\bibfnamefont
  {P.}~\bibnamefont {Sch{\"o}nfeldt}}, \bibinfo {author} {\bibfnamefont
  {M.}~\bibnamefont {Schuh}}, \bibinfo {author} {\bibfnamefont
  {J.}~\bibnamefont {Steinmann}}, \bibinfo {author} {\bibfnamefont
  {M.}~\bibnamefont {Yan}}, \bibinfo {author} {\bibfnamefont {E.}~\bibnamefont
  {Br{\"u}ndermann}}, \bibinfo {author} {\bibfnamefont {M.}~\bibnamefont
  {Weber}}, \bibinfo {author} {\bibfnamefont {A.-S.}\ \bibnamefont
  {M{\"u}ller}}, \bibinfo {author} {\bibfnamefont {G.}~\bibnamefont {Borghi}},
  \bibinfo {author} {\bibfnamefont {M.}~\bibnamefont {Boscardin}}, \ and\
  \bibinfo {author} {\bibfnamefont {S.}~\bibnamefont {Ronchin}},\ }in\
  \href@noop {} {\emph {\bibinfo {booktitle} {Proceedings of IBIC'17}}}\
  (\bibinfo {year} {2017})\ pp.\ \bibinfo {pages} {12--17}\BibitemShut
  {NoStop}%
\bibitem [{\citenamefont {Funkner}\ \emph {et~al.}()\citenamefont {Funkner},
  \citenamefont {Blomley}, \citenamefont {Br{\"{u}}ndermann}, \citenamefont
  {Caselle}, \citenamefont {Hiller}, \citenamefont {Nasse}, \citenamefont
  {Niehues}, \citenamefont {Rota}, \citenamefont {Sch{\"{o}}nfeldt},
  \citenamefont {Walther}, \citenamefont {Weber},\ and\ \citenamefont
  {M{\"{u}}ller}}]{Funkner2018}%
  \BibitemOpen
  \bibfield  {author} {\bibinfo {author} {\bibfnamefont {S.}~\bibnamefont
  {Funkner}}, \bibinfo {author} {\bibfnamefont {E.}~\bibnamefont {Blomley}},
  \bibinfo {author} {\bibfnamefont {E.}~\bibnamefont {Br{\"{u}}ndermann}},
  \bibinfo {author} {\bibfnamefont {M.}~\bibnamefont {Caselle}}, \bibinfo
  {author} {\bibfnamefont {N.}~\bibnamefont {Hiller}}, \bibinfo {author}
  {\bibfnamefont {M.~J.}\ \bibnamefont {Nasse}}, \bibinfo {author}
  {\bibfnamefont {G.}~\bibnamefont {Niehues}}, \bibinfo {author} {\bibfnamefont
  {L.}~\bibnamefont {Rota}}, \bibinfo {author} {\bibfnamefont {P.}~\bibnamefont
  {Sch{\"{o}}nfeldt}}, \bibinfo {author} {\bibfnamefont {S.}~\bibnamefont
  {Walther}}, \bibinfo {author} {\bibfnamefont {M.}~\bibnamefont {Weber}}, \
  and\ \bibinfo {author} {\bibfnamefont {A.-S.}\ \bibnamefont {M{\"{u}}ller}},\
  }\href@noop {} {\bibfield  {journal} {\bibinfo  {journal} {arXiv.org}\
  }}\Eprint {http://arxiv.org/abs/1809.07530} {arXiv:1809.07530} \BibitemShut
  {NoStop}%
\bibitem [{\citenamefont {Hofmann}\ \emph {et~al.}(2010)\citenamefont
  {Hofmann}, \citenamefont {Birkel}, \citenamefont {Fitterer}, \citenamefont
  {Hillenbrand}, \citenamefont {Hiller}, \citenamefont {Huttel}, \citenamefont
  {Judin}, \citenamefont {Klein}, \citenamefont {Marsching}, \citenamefont
  {M{\"u}ller}, \citenamefont {Smale}, \citenamefont {Sonnad},\ and\
  \citenamefont {Tavares}}]{Hofmann_IPAC_10}%
  \BibitemOpen
  \bibfield  {author} {\bibinfo {author} {\bibfnamefont {A.}~\bibnamefont
  {Hofmann}}, \bibinfo {author} {\bibfnamefont {I.}~\bibnamefont {Birkel}},
  \bibinfo {author} {\bibfnamefont {M.}~\bibnamefont {Fitterer}}, \bibinfo
  {author} {\bibfnamefont {S.}~\bibnamefont {Hillenbrand}}, \bibinfo {author}
  {\bibfnamefont {N.}~\bibnamefont {Hiller}}, \bibinfo {author} {\bibfnamefont
  {E.}~\bibnamefont {Huttel}}, \bibinfo {author} {\bibfnamefont
  {V.}~\bibnamefont {Judin}}, \bibinfo {author} {\bibfnamefont
  {M.}~\bibnamefont {Klein}}, \bibinfo {author} {\bibfnamefont
  {S.}~\bibnamefont {Marsching}}, \bibinfo {author} {\bibfnamefont {A.-S.}\
  \bibnamefont {M{\"u}ller}}, \bibinfo {author} {\bibfnamefont
  {N.}~\bibnamefont {Smale}}, \bibinfo {author} {\bibfnamefont
  {K.}~\bibnamefont {Sonnad}}, \ and\ \bibinfo {author} {\bibfnamefont
  {P.}~\bibnamefont {Tavares}},\ }in\ \href@noop {} {\emph {\bibinfo
  {booktitle} {Proceedings of IPAC'10}}}\ (\bibinfo {year} {2010})\ pp.\
  \bibinfo {pages} {924--926}\BibitemShut {NoStop}%
\bibitem [{\citenamefont {Kehrer}\ \emph {et~al.}(2016)\citenamefont {Kehrer},
  \citenamefont {Blomley}, \citenamefont {Brosi}, \citenamefont
  {Br{\"u}ndermann}, \citenamefont {Hiller}, \citenamefont {M{\"u}ller},
  \citenamefont {Nasse}, \citenamefont {Schedler}, \citenamefont
  {Sch{\"o}nfeldt}, \citenamefont {Schuh}, \citenamefont {Sch{\"u}tze},
  \citenamefont {Smale},\ and\ \citenamefont {Steinmann}}]{KehrerIPAC2016}%
  \BibitemOpen
  \bibfield  {author} {\bibinfo {author} {\bibfnamefont {B.}~\bibnamefont
  {Kehrer}}, \bibinfo {author} {\bibfnamefont {E.}~\bibnamefont {Blomley}},
  \bibinfo {author} {\bibfnamefont {M.}~\bibnamefont {Brosi}}, \bibinfo
  {author} {\bibfnamefont {E.}~\bibnamefont {Br{\"u}ndermann}}, \bibinfo
  {author} {\bibfnamefont {N.}~\bibnamefont {Hiller}}, \bibinfo {author}
  {\bibfnamefont {A.-S.}\ \bibnamefont {M{\"u}ller}}, \bibinfo {author}
  {\bibfnamefont {M.~J.}\ \bibnamefont {Nasse}}, \bibinfo {author}
  {\bibfnamefont {M.}~\bibnamefont {Schedler}}, \bibinfo {author}
  {\bibfnamefont {P.}~\bibnamefont {Sch{\"o}nfeldt}}, \bibinfo {author}
  {\bibfnamefont {M.}~\bibnamefont {Schuh}}, \bibinfo {author} {\bibfnamefont
  {P.}~\bibnamefont {Sch{\"u}tze}}, \bibinfo {author} {\bibfnamefont
  {N.}~\bibnamefont {Smale}}, \ and\ \bibinfo {author} {\bibfnamefont {J.~L.}\
  \bibnamefont {Steinmann}},\ }in\ \href {\doibase
  doi:10.18429/JACoW-IPAC2016-MOPMB014} {\emph {\bibinfo {booktitle}
  {Proceedings of IPAC'16}}}\ (\bibinfo {year} {2016})\ pp.\ \bibinfo {pages}
  {109--111}\BibitemShut {NoStop}%
\bibitem [{\citenamefont {Steinmann}\ \emph {et~al.}(2018)\citenamefont
  {Steinmann}, \citenamefont {Brosi}, \citenamefont {Br\"undermann},
  \citenamefont {Caselle}, \citenamefont {Funkner}, \citenamefont {Kehrer},
  \citenamefont {Nasse}, \citenamefont {Niehues}, \citenamefont {Rota},
  \citenamefont {Sch\"onfeldt}, \citenamefont {Siegel}, \citenamefont {Weber},\
  and\ \citenamefont {M\"uller}}]{Steinmann_FLS_2018}%
  \BibitemOpen
  \bibfield  {author} {\bibinfo {author} {\bibfnamefont {J.}~\bibnamefont
  {Steinmann}}, \bibinfo {author} {\bibfnamefont {M.}~\bibnamefont {Brosi}},
  \bibinfo {author} {\bibfnamefont {E.}~\bibnamefont {Br\"undermann}}, \bibinfo
  {author} {\bibfnamefont {M.}~\bibnamefont {Caselle}}, \bibinfo {author}
  {\bibfnamefont {S.}~\bibnamefont {Funkner}}, \bibinfo {author} {\bibfnamefont
  {B.}~\bibnamefont {Kehrer}}, \bibinfo {author} {\bibfnamefont
  {M.}~\bibnamefont {Nasse}}, \bibinfo {author} {\bibfnamefont
  {G.}~\bibnamefont {Niehues}}, \bibinfo {author} {\bibfnamefont
  {L.}~\bibnamefont {Rota}}, \bibinfo {author} {\bibfnamefont {P.}~\bibnamefont
  {Sch\"onfeldt}}, \bibinfo {author} {\bibfnamefont {M.}~\bibnamefont
  {Siegel}}, \bibinfo {author} {\bibfnamefont {M.}~\bibnamefont {Weber}}, \
  and\ \bibinfo {author} {\bibfnamefont {A.-S.}\ \bibnamefont {M\"uller}},\
  }in\ \href@noop {} {\emph {\bibinfo {booktitle} {Proceedings of the 60th ICFA
  Advanced Beam Dynamics Workshop on Future Light Sources}}}\ (\bibinfo {year}
  {2018})\BibitemShut {NoStop}%
\bibitem [{\citenamefont {Kehrer}\ \emph {et~al.}(2017)\citenamefont {Kehrer},
  \citenamefont {Blomley}, \citenamefont {Brosi}, \citenamefont
  {Br{\"u}ndermann}, \citenamefont {Hiller}, \citenamefont {M{\"u}ller},
  \citenamefont {Nasse}, \citenamefont {Schedler}, \citenamefont {Schuh},
  \citenamefont {Schwarz}, \citenamefont {Sch{\"o}nfeldt}, \citenamefont
  {Sch{\"u}tze}, \citenamefont {Smale},\ and\ \citenamefont
  {Steinmann}}]{kehrer2017time}%
  \BibitemOpen
  \bibfield  {author} {\bibinfo {author} {\bibfnamefont {B.}~\bibnamefont
  {Kehrer}}, \bibinfo {author} {\bibfnamefont {E.}~\bibnamefont {Blomley}},
  \bibinfo {author} {\bibfnamefont {M.}~\bibnamefont {Brosi}}, \bibinfo
  {author} {\bibfnamefont {E.}~\bibnamefont {Br{\"u}ndermann}}, \bibinfo
  {author} {\bibfnamefont {N.}~\bibnamefont {Hiller}}, \bibinfo {author}
  {\bibfnamefont {A.-S.}\ \bibnamefont {M{\"u}ller}}, \bibinfo {author}
  {\bibfnamefont {M.}~\bibnamefont {Nasse}}, \bibinfo {author} {\bibfnamefont
  {M.}~\bibnamefont {Schedler}}, \bibinfo {author} {\bibfnamefont
  {M.}~\bibnamefont {Schuh}}, \bibinfo {author} {\bibfnamefont
  {M.}~\bibnamefont {Schwarz}}, \bibinfo {author} {\bibfnamefont
  {P.}~\bibnamefont {Sch{\"o}nfeldt}}, \bibinfo {author} {\bibfnamefont
  {P.}~\bibnamefont {Sch{\"u}tze}}, \bibinfo {author} {\bibfnamefont
  {N.}~\bibnamefont {Smale}}, \ and\ \bibinfo {author} {\bibfnamefont
  {J.}~\bibnamefont {Steinmann}},\ }in\ \href {\doibase
  10.18429/JACoW-IPAC2017-MOOCB1} {\emph {\bibinfo {booktitle} {Proceedings of
  IPAC'17}}}\ (\bibinfo {year} {2017})\ pp.\ \bibinfo {pages}
  {53--56}\BibitemShut {NoStop}%
\bibitem [{\citenamefont {Sch\"onfeldt}\ \emph {et~al.}(2017)\citenamefont
  {Sch\"onfeldt}, \citenamefont {Brosi}, \citenamefont {Schwarz}, \citenamefont
  {Steinmann},\ and\ \citenamefont {M\"uller}}]{PRAB_Inovesa}%
  \BibitemOpen
  \bibfield  {author} {\bibinfo {author} {\bibfnamefont {P.}~\bibnamefont
  {Sch\"onfeldt}}, \bibinfo {author} {\bibfnamefont {M.}~\bibnamefont {Brosi}},
  \bibinfo {author} {\bibfnamefont {M.}~\bibnamefont {Schwarz}}, \bibinfo
  {author} {\bibfnamefont {J.~L.}\ \bibnamefont {Steinmann}}, \ and\ \bibinfo
  {author} {\bibfnamefont {A.-S.}\ \bibnamefont {M\"uller}},\ }\href {\doibase
  10.1103/PhysRevAccelBeams.20.030704} {\bibfield  {journal} {\bibinfo
  {journal} {Phys. Rev. Accel. Beams}\ }\textbf {\bibinfo {volume} {20}},\
  \bibinfo {pages} {030704} (\bibinfo {year} {2017})}\BibitemShut {NoStop}%
\bibitem [{\citenamefont {Boltz}(2017)}]{Boltz_master}%
  \BibitemOpen
  \bibfield  {author} {\bibinfo {author} {\bibfnamefont {T.}~\bibnamefont
  {Boltz}},\ }\emph {\bibinfo {title} {{Comprehensive Analysis of
  Micro-Structure Dynamics in Longitudinal Electron Bunch Profiles}}},\ \href
  {\doibase 10.5445/IR/1000068253} {\bibinfo {type} {Master thesis}},\ \bibinfo
   {school} {Karlsruhe Institute of Technology} (\bibinfo {year}
  {2017})\BibitemShut {NoStop}%
\end{thebibliography}%

\end{document}